\documentclass[12pt,preprint]{aastex}

\usepackage{natbib}                % To format bibliographies.
\usepackage{verbatim}
\usepackage{subfigure}
\usepackage{epsfig}

\def\arcsec{\ifmmode '' \else $''$\fi}

\def\arcsecpoint{\ifmmode ''\!. \else $''\!.$\fi}

\def\kms{\ifmmode {\rm km\ s}^{-1} \else km s$^{-1}$\fi}
\def\Msun{\ifmmode {\rm M}_{\odot} \else M$_{\odot}$\fi}
\def\Lsun{\ifmmode {\rm L}_{\odot} \else L$_{\odot}$\fi}
\def\Zsun{\ifmmode {\rm Z}_{\odot} \else Z$_{\odot}$\fi}

\def\ergscm2{ergs\,s$^{-1}$\,cm$^{-2}$}
\def\icm3{{\rm cm}^{-3}}
\def\icm2{{\rm cm}^{-2}}
\def\qo{\ifmmode q_{\rm o} \else $q_{\rm o}$\fi}
\def\Ho{\ifmmode H_{\rm o} \else $H_{\rm o}$\fi}
\def\ho{\ifmmode h_{\rm o} \else $h_{\rm o}$\fi}

\def\vFWHM{\ifmmode v_{\mbox{\tiny FWHM}} \else
            $v_{\mbox{\tiny FWHM}}$\fi}
\def\CCF{\ifmmode F_{\it CCF} \else $F_{\it CCF}$\fi}
\def\ACF{\ifmmode F_{\it ACF} \else $F_{\it ACF}$\fi}
\def\Halpha{\ifmmode {\rm H}\alpha \else H$\alpha$\fi}
\def\Hbeta{\ifmmode {\rm H}\beta \else H$\beta$\fi}
\def\Hgamma{\ifmmode {\rm H}\gamma \else H$\gamma$\fi}
\def\Hdelta{\ifmmode {\rm H}\delta \else H$\delta$\fi}
\def\Lya{\ifmmode {\rm Ly}\alpha \else Ly$\alpha$\fi}
\def\Lyb{\ifmmode {\rm Ly}\beta \else Ly$\beta$\fi}
\def\Lyg{\ifmmode {\rm Ly}\beta \else Ly$\gamma$\fi}
\def\fei{Fe\,{\sc i}}
\def\feii{Fe\,{\sc ii}}
\def\feiii{Fe\,{\sc iii}}

\def\hi{H\,{\sc i}}

\def\hei{He\,{\sc i}} 
\def\heii{He\,{\sc ii}}
\def\ci{C\,{\sc i}}
\def\cii{C\,{\sc ii}}
\def\ciii{\ifmmode {\rm C}\,{\sc iii} \else C\,{\sc iii}\fi}
\def\civ{\ifmmode {\rm C}\,{\sc iv} \else C\,{\sc iv}\fi}
\def\cv{\ifmmode {\rm C}\,{\sc v} \else C\,{\sc v}\fi}
\def\cvi{\ifmmode {\rm C}\,{\sc vi} \else C\,{\sc vi}\fi}

\def\oi{O\,{\sc i}}

\def\o5007{[O\,{\sc iii}]\,$\lambda5007$}

\def\mgi{Mg\,{\sc i}}
\def\mnii{Mn\,{\sc ii}}

\def\mgii{Mg\,{\sc ii}}

\def\siiv{Si\,{\sc iv}}

\def\siII{Si\,{\sc ii}}

\def\niII{Ni\,{\sc ii}}

\def\siv{S\,{\sc iv}}

\def\fei{Fe\,{\sc i}}
\def\feii{Fe\,{\sc ii}}
\def\feiii{Fe\,{\sc iii}}
\def\alii{Al\,{\sc ii}}
\def\aliii{Al\,{\sc iii}}

\def\coii{Co\,{\sc ii}}
\def\crii{Cr\,{\sc ii}}
\def\znii{Zn\,{\sc ii}}
\def\o{\o}

\newcommand{\vy}[2]{#1_{\scriptscriptstyle #2}}
\def\gtorder{\mathrel{\raise.3ex\hbox{$>$}\mkern-14mu
             \lower0.6ex\hbox{$\sim$}}}
\def\ltorder{\mathrel{\raise.3ex\hbox{$<$}\mkern-14mu
             \lower0.6ex\hbox{$\sim$}}}
\def\proptwid{\mathrel{\raise.3ex\hbox{$\propto$}\mkern-14mu
             \lower0.6ex\hbox{$\sim$}}}

\begin{document}

\shortauthors{Dunn, et al.}
\shorttitle{The Multi-component Absorber QSO 0318-0600}

\title{The Quasar Outflow Contribution to AGN Feedback: VLT Measurements of SDSS~J0318-0600\footnote{Based on observations collected at the European Organisation for Astronomical Research in the Southern Hemisphere, Chile; under program 078.B-0433.}}

%\title{Kinetic Luminosity Determination of Quasar Outflow in SDSS~J0318-0600}

\author{Jay P. Dunn\altaffilmark{1},
Manuel Bautista\altaffilmark{1,2},
Nahum Arav\altaffilmark{1},
Max Moe\altaffilmark{3,4},
Kirk Korista\altaffilmark{2},
Elisa Costantini\altaffilmark{5},
Chris Benn\altaffilmark{6},
Sara Ellison\altaffilmark{7},
\& Doug Edmonds\altaffilmark{1}}

\altaffiltext{1}{Department of Physics, Virginia Tech, Blacksburg, VA 24061.
Email: jdunn77@vt.edu, arav@vt.edu, edmonds@vt.edu}
\altaffiltext{2}{Department of Physics, Western Michigan University, Kalamazoo, MI 49008-5252 Email: kirk.korista@wmich.edu, manuel.bautista@wmich.edu}
\altaffiltext{3}{Department of Astrophysical and Planetary Sciences and Center for Astrophysics and Space Astronomy, University of Colorado, 389-UCB, Boulder, CO 80309.}
\altaffiltext{4}{Harvard-Smithsonian Center for Astrophysics, Cambridge, MA 02138 Email: mmoe@cfa.harvard.edu}
\altaffiltext{5}{SRON National Institute for Space Research, Sorbonnelaan 2, 3584 CA Utrecht, Netherlands}
\altaffiltext{6}{Isaac Newton Group, Apartado 321, E-38700 Santa Cruz de La Palma, Spain}
\altaffiltext{7}{Department of Physics and Astronomy, University of Victoria, Victoria, BC V8P 1A1, Canada}

\begin{abstract}

We present high spectral resolution VLT observations of the BAL quasar 
SDSS~J0318-0600. This high quality data set allows us to extract accurate 
ionic column densities and determine an electron number density of 
$n_e$=10$^{3.3\pm0.2}$ cm$^{-3}$ for the main outflow absorption component. 
The heavily reddened spectrum of SDSS~J0318-0600 requires purely silicate 
dust with a reddening curve characteristic of predominately large grains, 
from which we estimate the bolometric luminosity. We carry out 
photoionization modeling to determine the total column density, ionization 
parameter and distance of the gas and find that the photionization models 
suggest abundances greater than solar. Due to the uncertainty in the 
location of the dust extinction, we arrive at two viable distances for 
the main ouflow component from the central source, 6 and 18 kpc, where 
we consider the 6 kpc location as somewhat more physically plausable. 
Assuming the canonical global covering of 20\% for the outflow and a distance
of 6 kpc, our analysis yields a mass flux of 120 M$_{\sun}$ yr$^{-1}$ and a 
kinetic luminosity that is $\sim$0.1\% of the bolometric luminosity of the 
object. Should the dust be part of the outflow, then these values are 
$\sim$4$\times$ larger. The large mass flux and kinetic luminosity make 
this outflow a significant contributor to AGN feedback processes.

\end{abstract}

%\keywords{galaxies: individual (SDSS J0318-0600); line: formation; quasars: absorption lines}
%\keywords{quasars: individual (SDSS J0318-0600); quasars: absorption lines}

\section{Introduction}

Quasar outflows, seen as ultraviolet blueshifted absorption features in 
their spectra, may contribute to a variety of AGN feedback mechanisms
\citep{2006MmSAI..77..573E}. Such feedback processes are invoked to
explain the: evolution of the super massive black hole (SMBH)
\citep{1998A&A...331L...1S, 1999MNRAS.303L...1B}; evolution of the
host galaxy (Di Matteo et al.  2005), shape of the observed quasar
luminosity function \citep{2005ApJ...630..705H, 2005MNRAS.361..776S,
2006ApJ...647..753M} and the enrichment of the
intra-cluster and inter-galactic media \citep{2007A&A...463..513M}.

The contribution of quasar outflows to AGN feedback depends mainly on
their mass flux ($\dot{M}$) and kinetic luminosity ($\dot{E}$$_k$),
both of which depend linearly on the the total hydrogen column
density (N$_{H}$) of the outflow and its characteristic distance ($R$) 
from the central source (see Equations 9 and 11 in $\S$~6). Until
recently, only loosely constrained estimates have been available for
N$_{H}$ and $R$ in the same outflow. \citet{1986ApJ...310...40M} and
\citet{1996ApJS..102..239T} were only able to find lower 
limits for the distance to outflows in GC~1556+335 and HS~1946+7658, 
respectively. \citet{1995ApJ...443..586W}, 
analyzing the spectrum of QSO~0059$-$2735, constrained the distance 
of the outflow to a range spanning over an order of 
magnitude. \citet{2001ApJ...548..609D} determined the distance to the
outflow observed in FIRST~J1044+3656 to within a factor of 100, and
had a similar uncertainty in the inferred N$_H$ (see their table 3), 
which led to an uncertainty in $\dot{M}$ and $\dot{E}$$_k$ of 
approximately one order of magnitude. In the cases of FIRST~J1214+2803
\& FIRST~J0840+3633, \citet{2002ApJ...570..514D, 2002ApJ...567...58D}
were able to constrain $\dot{M}$ and $\dot{E}$$_k$ to a similar
accuracy. For the quasar 3C~191, \citet{2001ApJ...550..142H} 
estimated $\dot{M}$ and $\dot{E}$$_k$ to within a factor of a few. 
However, beside their internal uncertainties, all the works cited above 
suffered from unquantified systematic errors due
to the use of inadequate absorption models that led to unreliable
measurements of column densities in the observed troughs, which are
crucial for determining almost every physical aspect of the outflows:
the ionization equilibrium and abundances, number density, distance,
mass flux, and kinetic luminosity.

In order to determine accurate $\dot{M}$ and $\dot{E}$$_k$ in quasar
outflows we launched a multistage research program: We first developed
analysis techniques that allow us to determine reliable trough column
densities \citep{1999ApJ...524..566A,1999ApJ...516...27A,
2002ApJ...566..699A,2005ApJ...620..665A,2008ApJ...681..954A,
2004ApJS..152....1S,2005ApJ...623...85G}. This was followed by identifying
outflow targets in which we can find diagnostics for both N$_{H}$ and
$R$.  We searched through $\sim$50,000 QSOs with z$>$0.25 and r$<$19.3
in the SDSS Data Release~6 \citep{2008ApJS..175..297A}, $\sim$100
objects were found to exhibit intrinsic outflow absorption systems
containing troughs from excited or metastable levels of one or more of
the following ions: \hei, \siII, \feii, \feiii\ or \niII.  The key
spectral features to obtain distances and total column densities for
outflows are a) absorption troughs of excited and metastable levels from 
which we can determine the electron density, and b) the availability of 
troughs from
many ions, including lines from multiple ions of the same element. We
then began observing these objects with a combination of high spectral
resolution, large spectral coverage and high signal-to-noise ratios
that in aggregate allows for the use of the analysis techniques
mentioned above. \citet{2008ApJ...681..954A} and 
\citet{2008arXiv0807.0230K} used
this analysis method on VLT observations of QSO~2359--1241, deriving
N$_{H}$ and hydrogen number density determinations with accuracies
better than 25\% and 20\%, respectively.

The object we analyze in this paper, SDSS~J0318$-$0600 was found in
our search through the SDSS data base 
\citep[although first discovered by][]{2002ApJS..141..267H}. 
We chose this object for a high resolution follow up observation, 
due to its high apparent brightness and its suitability for VLT 
observations. SDSS~J0318$-$0600 is a luminous quasar 
\citep[log L$_{Bol}$$\approx$47.7 ergs s$^{-1}$, see $\S$5; z=1.9668;][]
{2002ApJS..141..267H,2003AJ....126.2579S} that shows a myriad of outflow 
components ranging from a high ionization broad absorption
line system to several narrow absorption line (NAL) features spanning
a velocity range from $-$2800 to $-$7500 km s$^{-1}$, with respect
to the quasar's rest frame. In a survey by Hall et al. (2002), it was
classified as an iron low-ionization broad absorption line system
(FeLoBAL) and listed as an unusual quasar due to
the BAL located at $\sim$27,000 km s$^{-1}$ and the system of NALs,
which appears as a BAL in higher ionization lines such as \civ. They
noted that the \feii\ UV1, UV2 and UV3 multiplets were detected in
absorption and that the UV38 multiplet appeared to be present as
well. It was also noted that SDSS~J0318$-$0600 has a heavily reddened 
spectrum. \citet{2006ApJ...639..766P} listed this object 
as a strong \mgii\ absorber with a rest frame absorption width $>$1.4 
\AA, in a search of \mgii\ absorbing objects.  \citet{2006PASP..118.1077H}
showed that the object's absorption was strong, particularly in \siII\
$\lambda$ 1808, and labeled the absorber as a damped Lyman alpha system
(DLA). However, in our VLT data, we detect absorption from excited
ionic energy levels that conclusively classify the system as an
outflow \citep{1997ApJ...478...80H}.

The outline for this paper is as follows, in $\S$ 2, we present the
VLT eschelle observations and their reduction. For $\S$ 3,
we discuss our methods of determining and measuring individual
kinematic components. We then concentrate on a physical analysis
of the main component of the outflow: in section 4, we present our
photoionization analysis of the spectrum and determine the hydrogen column
density (N$_H$), electron density (n$_e$), distance (R) and temperature
(T) for the main absorption component. We derive the mass flux and
kinetic luminosity in section 5, and in $\S$~6 we discuss our results.

\begin{deluxetable}{lcccccc}
\tablewidth{0.7\textwidth}
\tablecolumns{7}
\footnotesize
\tablecaption{VLT Observations of QSO J0318$-$0600}

\tablehead{
\colhead{Date} &
\colhead{Time$^a$} &
\colhead{JD$^b$} &
\colhead{Airmass} &
\colhead{Seeing} \\
}

\startdata
30 Sep 2006 & 04:35:01 & 54008.19092 & 1.41 & 1.08 \\
01 Oct 2006 & 03:51:30 & 54009.25268 & 1.12 & 1.07 \\
01 Oct 2006 & 06:53:51 & 54009.28740 & 1.06 & 0.97 \\
15 Oct 2006 & 05:55:23 & 54023.24679 & 1.06 & 1.00 \\
19 Oct 2006 & 05:34:22 & 54027.23220 & 1.07 & 0.82 \\
20 Oct 2006 & 06:49:21 & 54028.28427 & 1.07 & 0.71 \\
22 Oct 2006 & 04:45:58 & 54030.20063 & 1.10 & 0.66 \\
22 Oct 2006 & 05:38:28 & 54030.23505 & 1.06 & 0.55 \\
\enddata
\normalsize
\tablenotetext{a}{Exposure time for each observation was 2800s}
\tablenotetext{b}{Julian Date presented in days +2400000}
\label{obs}
\end{deluxetable}
\clearpage

During October of 2006, we obtained 8 echelle spectra, each with an
exposure time of 2850 sec, of the object SDSS~J0318$-$0600
\citep[z=1.967,][RA 03:18:56.7, Dec $-$06:00:37.4]{2007AJ....134..102S}
with the VLT. The observations were carried out by European Southern
Observatory (ESO) staff and we present the observational parameters in
Table~\ref{obs}. The spectra were taken using the UV-Visual Echelle Spectrograph
(UVES) with a slit width of 1.0 arcsec, and a central wavelength position
of 5800 and 3460 \AA\ for the red (Grating/Filter=CD1/HER5) and blue
(Grating/Filter=CD3/SHP700) detectors respectively. A slit width of 1.0
arcsec gives a resolving power of $\sim$40000. The data were
read out in a 2$\times$2 pixel-binned fashion, in high gain mode.

The data were extracted and calibrated using the standard UVES data
reduction pipeline, based on ECHELLE routines, which is part of the
reduction package MIDAS \citep{2000SPIE.4010..246B}. This calibration
process can introduce quasi-periodic oscillations in amplitude in the
final spectral product \citep[see][on their analysis of QSO~2359-1241]
{2008ApJ...681..954A}, which can create up to 2-3\% systematic errors in
continuum placement. The observations were corrected to heliocentric and
subject to a cosmic ray removal procedure ($\sigma$ $>$ 3.5). We
re-binned the data to 7.0 km s$^{-1}$, approximately one resolution
element, and normalized the spectrum using the methods described in
\citet{2008ApJ...681..954A}.

\section{Detection \& Measurements}

\begin{figure}[!h]
  \centering \includegraphics[angle=90,width=0.7\textwidth]
  {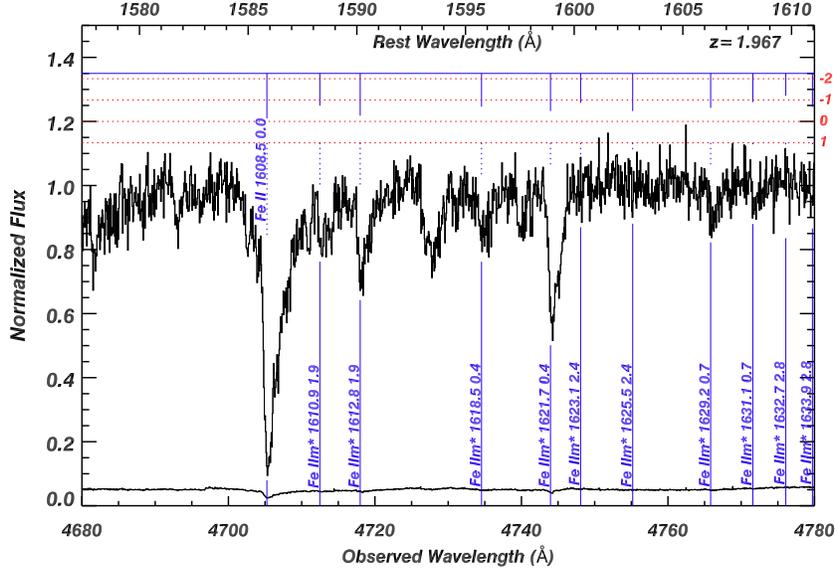}\\
  \caption[]
  {\footnotesize
Portion of the VLT spectrum showing the \feii\ UV 38 multiplet
in SDSS~J0318$-$0600 (the full spectrum can be found online). Line
identifications are indicated by vertical blue lines for the expected
position of component $\bf i$, based on the redshift z=1.9257.
The redshift of the quasar is in black located at the top right. The
log($gf$) values of the identified transitions are shown above the
spectrum, where $g$ is the statistical weight and $f$ is the oscillator
strength. The size of the vertical blue line provides the value, which
can be read from the y-axis scale at the top right. Each identifier
lists the ion, wavelength, and lower energy level (in 1000s of
cm$^{-1}$). The spectrum error is plotted in solid black below the
spectrum ($\approx$0.05 in normalized flux).}
  \label{f1}
\end{figure}

The spectrum of SDSS~J0318$-$0600 contains a plethora of absorption 
features available for potential diagnostics. In Figure \ref{f1}, we 
show a sample of the VLT spectrum with line identifications, and 
present the full spectrum with identifications in our online figure. 
These identifications come from matching the \mgii\ doublet lines 
and plotting location identifiers for other ionic lines. We also 
provide an online table for all identified lines, their wavelengths, 
oscillator strengths and references. We show in Figure 1 a portion of
the full online figure that contains readily visible lines from both 
the ground as well as excited levels of \feii\ (lower energy levels 
up to 1873 cm$^{-1}$).

\subsection{Identification of Kinematic Components}

We identify each of the individual kinematic absorption components in 
the spectrum using the \alii\ $\lambda$1671 absorption line profile, 
which is a singlet transition with no nearby strong absorption features. 
In Figure \ref{f2}, we show that there are approximately 11 separate 
kinematic components (labeled $\bf a-k$), with component $\bf e$ 
clearly a blend of multiple inseparable components. Using \alii\ 
as a template for the absorbers we search other ionic species for each 
of the 11 components.

Next, we identify the broad, strong $\bf i$ component in as many
lines as possible and select unblended troughs from which we are
able to measure ionic column densities. Figure \ref{f1} shows the
UV 38 \feii\ multiplet, which exhibits several absorption features
from component $\bf i$. We note that the absorption feature at
$\sim$4727 \AA\ is the $\bf k$ component of the \feii\ $\lambda$1608
resonance line. While this identification scheme is sufficient to
locate the majority of the lines that show an $\bf i$ component, we
also use the \alii\ as a template (see $\S$3.3), to search for the
other 10 components in the \feii\ UV 1, 2 \& 3 multiplet absorption
complexes.

%\clearpage
\noindent
\begin{rotate}
\begin{figure}
 \noindent
 \hspace{-2.5in}
 \begin{minipage}[t]{4.0cm}%\linewidth} A minipage that covers half the page
  \vspace{-2.5cm}
  \includegraphics[width=13.5cm,height=17.5cm]{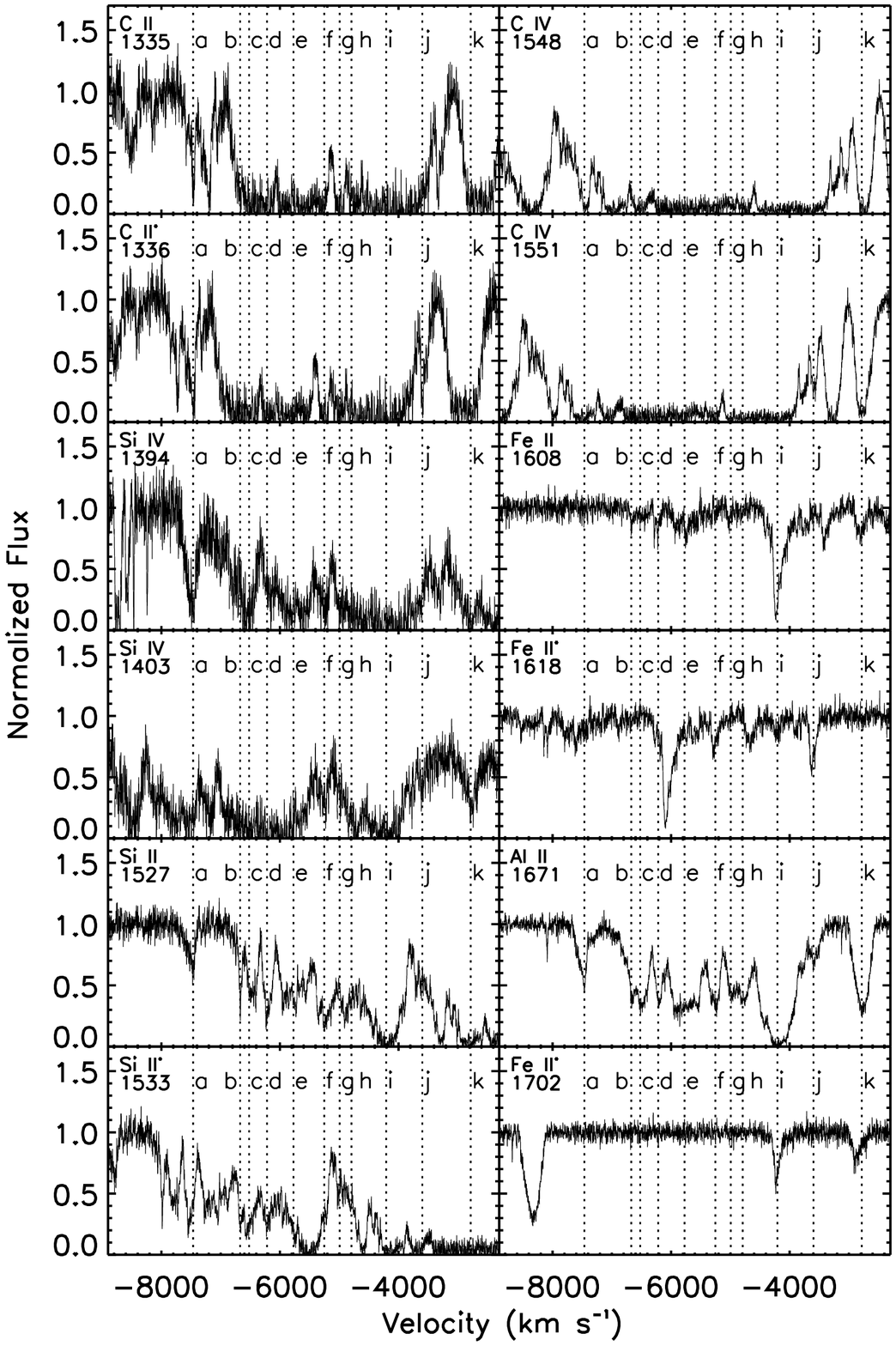}
 \end{minipage}
 \hspace{3.2in} % To get a little bit of space between the figures
 \begin{minipage}[t]{4.0cm}%\linewidth}
  \vspace{-2.5cm}
  \includegraphics[width=13.5cm,height=17.5cm]{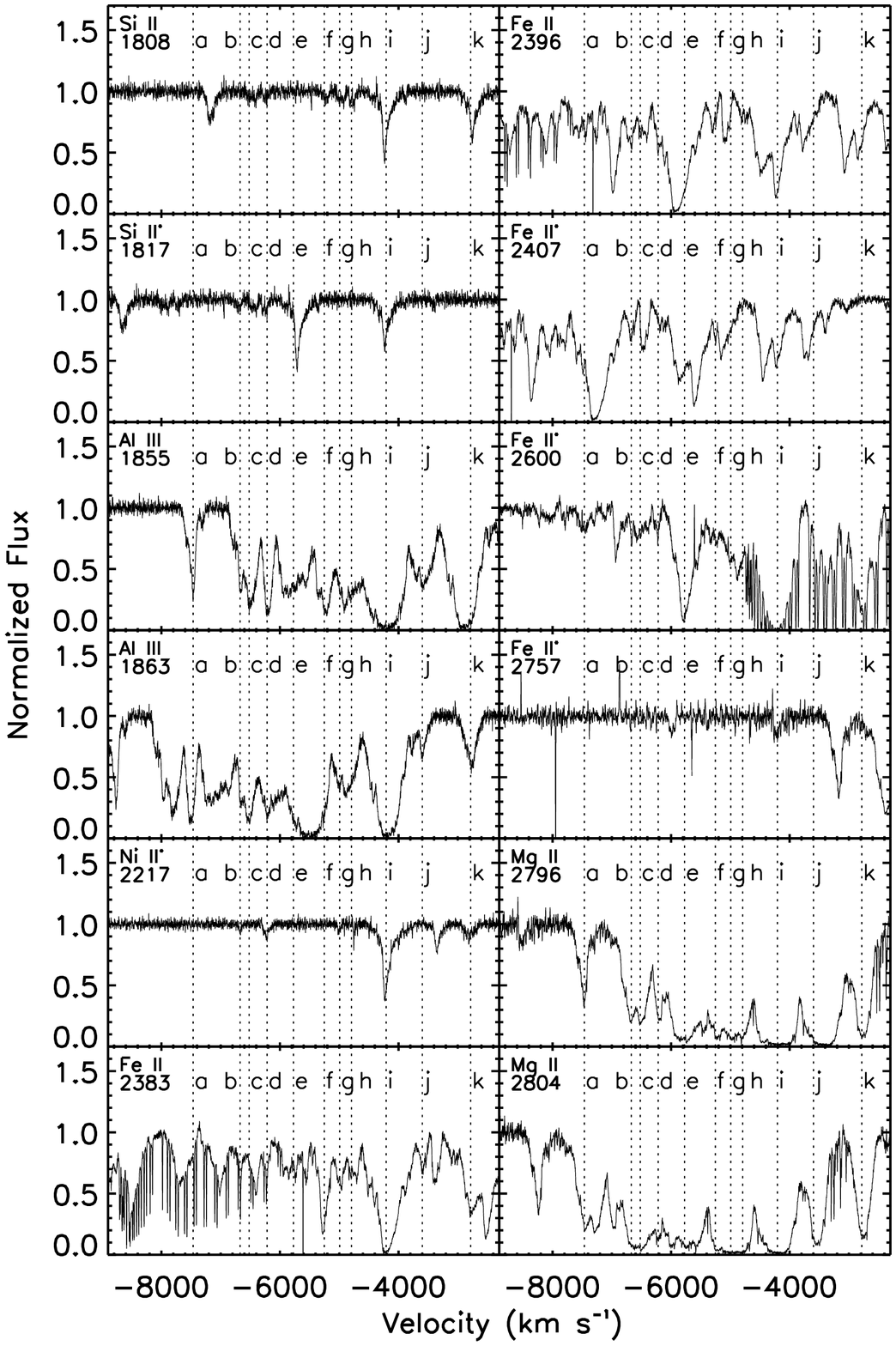}
 \end{minipage}
\caption{\scriptsize The main outflow troughs in SDSS~J0318$-$0600, 
ordered by wavelength. We identify the 11 kinematic components found in
\alii\ from high to low velocities, which we label $\bf a$ through $\bf k$.
In the plots of \feii\ $\lambda$2383, $\lambda$2396, and $\lambda$2600 there
are several contaminate lines present due to the telluric A and B bands as 
well as blending from other \feii\ troughs (identified in online figure). 
Each doublet also shows blending between components (e.g. \civ, \mgii).}
\label{f2}
\end{figure}
\end{rotate}

%\clearpage
%\newpage
\begin{deluxetable}{lrcrrcc}
\tablecolumns{7}
%\tablewidth{1.2\textwidth}
\tabletypesize{\footnotesize}
\tablecaption{Component $\bf i$: Measured Column Densities}
\tablehead{
\colhead{Ion} &
\colhead{Energy Level} &
\colhead{Velocity Range} & 
\colhead{Column Density} &
\colhead{Uncertainty$^a$} &
\colhead{Number} &
\colhead{Methods$^b$}\\

\colhead{ } &
\colhead{(E$_{low}$ cm$^{-2}$)} &
\colhead{(km s$^{-1}$)} &
\colhead{($\times$10$^{12}$ cm$^{-2}$)} &
\colhead{($\times$10$^{12}$ cm$^{-2}$)} &
\colhead{of Lines$^c$} & 
\colhead{ } \\
}

\startdata

\hei\   & 159856 & $-$4600 $-$3716 & $<$90       & ---  & 1 & 6 \\
\ci\    & 0      & $-$4600 $-$3716 & $<$37       & ---  & 1 & 6 \\
\cii\   & 63     & $-$4600 $-$3714 & $>$19000    & ---  & 1 & 4 \\
\civ\   & 0      & $-$4600 $-$3714 & 29000       & 3000 & 1 & 4 \\
\oi\    & 0      & $-$4551 $-$3896 & 1900        & $^{+2000}_{-1200}$ & 1 & 5 \\
\mgi\   & 0      & $-$4600 $-$3716 & $<$2        & ---  & 1 & 6 \\
\mgii\  & 0      & $-$4600 $-$3714 & 3200        & 400  & 1 & 4 \\
\alii\  & 0      & $-$4477 $-$3924 & 400         & 40   & 1 & 2 \\
\aliii\ & 0      & $-$4610 $-$3833 & 1560        & 220  & 1 & 3 \\
\siII\  & 0      & $-$4295 $-$4106 & 7220        & 720  & 1 & 1 \\
\siII\  & 287    & $-$4295 $-$4106 & 7380        & 740  & 1 & 1 \\
\siiv\  & 0      & $-$4600 $-$3714 & 5600        & 1300 & 1 & 4 \\
\crii\  & 0      & $-$4295 $-$4106 & 14.6        & 1.5  & 2 & 1 \\
\mnii\  & 0      & $-$4295 $-$4106 & 17.5        & 1.8  & 1 & 1 \\
\fei\   & 0      & $-$4600 $-$3714 & $<$3        & ---  & 1 & 6 \\
\feii\  & 0      & $-$4295 $-$4106 & 1275        & 128  & 3 & 1 \\
\feii\  & 385    & $-$4295 $-$4106 & 294         & 77   & 3 & 1 \\
\feii\  & 668    & $-$4295 $-$4106 & 64$^d$      & ---  & 2 & 1 \\
\feii\  & 863    & $-$4295 $-$4106 & 147$^d$     & ---  & 2 & 1 \\
\feii\  & 977    & $-$4295 $-$4106 & 31$^d$      & ---  & 1 & 1 \\
\feii\  & 1873   & $-$4295 $-$4106 & 163         & 16   & 4 & 1 \\
\feii\  & 2430   & $-$4295 $-$4106 & 25          & 5.4  & 1 & 1 \\
\feii\  & 7955   & $-$4295 $-$4106 & 8.1         & 0.8  & 3 & 1 \\
\coii\  & 3351   & $-$4295 $-$4106 & 9.4         & 0.9  & 1 & 1 \\
\niII\  & 0      & $-$4295 $-$4106 & 180         & 18   & 3 & 1 \\
\niII\  & 8394   & $-$4295 $-$4106 & 64          & 6.4  & 3 & 1 \\
\znii\  & 0      & $-$4295 $-$4106 & 8.4         & 0.8  & 2 & 1 \\

\enddata

\tablenotetext{a}{Reflects the statistical errors from line fitting. 
This does not reflect the reliability of the measurement methods, 
which have increasing systematic error in order of increasing label
number (see section 3.2).}

\tablenotetext{b}{Method used to determine column density, described 
by the respective label number; 1: Power Law fitting, 2: Velocity 
Dependent Covering Factor fitting, 3: Apparent optical depth fitting, 
4: ``wing'' fitting, 5: ``tip'' fitting, 6: Upper limit based on the 
noise method, see sections 3.2 and 3.3 for full details}

\tablenotetext{c}{Number of lines used in fitting. See Online Table 
for specific transition information.}

\tablenotetext{d}{Unreliable measurements due to weak lines}

\normalsize
\label{compi}
\end{deluxetable}
\thispagestyle{empty}

\clearpage

The \feii\ multiplets UV 1, 2 \& 3 are the best candidates to identify \feii\
lines from the other 10 components to measure, because they contain the 
\feii\ transitions with the strongest oscillator strengths. However, this 
task proves to be non-trivial in SDSS~J0318-0600. Identifying measurable 
lines from a saturated and heavily blended region (see Figure \ref{f2}) 
is difficult, especially when considering the large number of strong 
resonance and excited state lines. Our strategy is to make what line 
identifications for all 11 components that we can using our \alii\ 
template and search for other lines where the blending is not so 
prevalent. We provide the full list of detected ions, transition 
wavelengths, their oscillator strengths and lower level energies in the 
online table. 

We find that for the lowest and highest velocity components 
($\bf k$ and $\bf a$), enough measurements or limit determinations are 
available that we can effectively determine the electron density ($n_H$), 
the ionization parameter (U$_H$), and the total hydrogen column density 
($N_H$, see $\S$4). Unfortunately, blending prevents us from identifying 
enough measurable troughs for components $\bf b-h$ for proper analysis. 
We opt for the limiting case by taking the sum of these latter systems 
(see $\S$3.2).

\subsection{Column Density Determinations}

With a list of measurable absorption troughs, we 
extract ionic column densities. We begin the process by measuring 
lines from the strongest component, ($\bf i$), where we employ three
primary methods for determining the column densities. We describe 
three additional methods for column density determination that are less
reliable in section 3.3. We provide the measured values of ionic 
column densities as well as the method applied in Table~\ref{compi}.

Our first method of measuring column densities is to model a set of 
observed \feii\ resonance lines (e.g., Figure \ref{f3}) using an
inhomogenous absorption model, where we approximate the gas 
distribution across our line of sight to the background source with a 
power law \citep[see][especially Fig. 4, for a full description]
{2002ApJ...580...54D,2008ApJ...681..954A}. The optical depth at a 
given velocity within the outflow is described by: 
\begin{equation}
\tau_v(x) = \tau_{max}(v) x^a ,
\end{equation}
where $x$ is the spatial dimension in the plane of the sky 
\citep[simplified to one dimension with no loss of generality, see][]
{2005ApJ...620..665A}, $a$ is the power law distribution index, and
$\tau_{max}(v)$ is the highest value of $\tau$ at a given velocity. 
We simultaneously solve for 
both $\tau$$_{max}$ and $a$ for each velocity resolution element of 
the resonance \feii\ lines and apply this distribution ($a$) to the 
excited \feii\ lines. We then apply the same distribution dictated by 
the \feii\ lines to all ions because this is the simplest assumption 
that imposes the strongest constraints.

We convert the modeled $\tau$$_v$ to column density via 
\citep{1991ApJ...379..245S}:
\begin{equation}
N(v) = 3.8 \times 10^{14} \frac{1}{f\lambda} \tau_v~(cm^{-2} km^{-1} s),
\end{equation}
where $\lambda$ is the wavelength of the line in \AA, $f$ is the oscillator
strength, and $\tau$$_v$ is the optical depth in velocity space in 
km~s$^{-1}$, which is integrated over the spatial dimension $x$. Using 
Equation 2, we compute the column density for each 
velocity element and integrate over the range in velocity, which we justify 
later in this section. In Table~\ref{compi}, we 
quote the power law estimation of the column densities (when available) 
as this is the most accurate and most plausibly physical explanation of 
non-black saturated absorption at large distances 
\citep{2008ApJ...681..954A,2002ApJ...580...54D}. In Table~\ref{compi} and 
throughout this paper, we will refer to this as method 1.

Method 2 is to model the \feii\ resonance lines with a velocity 
dependent line of sight (LOS) covering factor. The LOS covering factor 
in this case is a geometric step-function distribution \citep[see 
Figure 4 in][]{2008ApJ...681..954A}. In other words, a fraction of the 
extended source is completely obscured by the outflow, while the rest 
of the source is uncovered. We assume that each radial velocity element 
has an independent LOS covering factor, which we again determine using 
the \feii\ resonance lines. We solve a fitting equation for $\tau$$_v$ 
\citep{2005ApJ...620..665A}:
\begin{equation}
I(v) = 1 - C(v) + C(v) e^{-\tau(v)},
\end{equation}
where $I$ is the normalized intensity within the trough and $C(v)$
is the velocity dependent LOS covering factor. First, we solve Equation
(3) for $C(v)$ and $\tau$$(v)$ using at least two resonance lines from 
\feii. We then calculate the velocity dependent column density using
$\tau$$(v)$ in Equation (2) and multiply the column density at each
velocity by $C(v)$
to obtain the average column density across the line of sight. Finally,
we compute the total ionic column density by integrating across the 
velocity range of the trough. This method is comparable in accuracy 
to method 1 \citep{2008ApJ...681..954A}, and the column density 
results are approximately the same as found by method 1.

The last of the three methods, method 3, is simply the apparent optical 
depth method \citep[AOD,][]{1991ApJ...379..245S}. 
This method assumes that all photons traveling to the observer pass
through the same amount of gas at a given velocity, which gives the
equation for normalized intensity: $I(v)$=$e$$^{-\tau(v)}$. We again 
convert to column density via Equation (2) and integrate to find the 
total ionic column density. This method does not account for partial LOS
covering and saturation effects in line centers, which can underestimate 
column densities by up to an order of magnitude. However, we find that 
for SDSS~J0318-066, in most cases a LOS covering factor of 1 (i.e. the 
AOD approximation) yields similar N$_{ion}$ values as those derived by 
the power law or partial covering method, a point 
we will discuss later in this section.

For the upcoming photoionization modeling, we adopt the ionic column 
densities as determined by method 1 (power law method) with some 
exceptions to this due to line blending that are noted in Table~\ref{compi}. 
We now present our analysis for the individual components.

\vspace{-0.7cm}
\subsubsection{Measurements of Component $\bf i$}

\begin{figure}[!h]
  \centering \includegraphics[angle=90,width=0.7\textwidth]
  {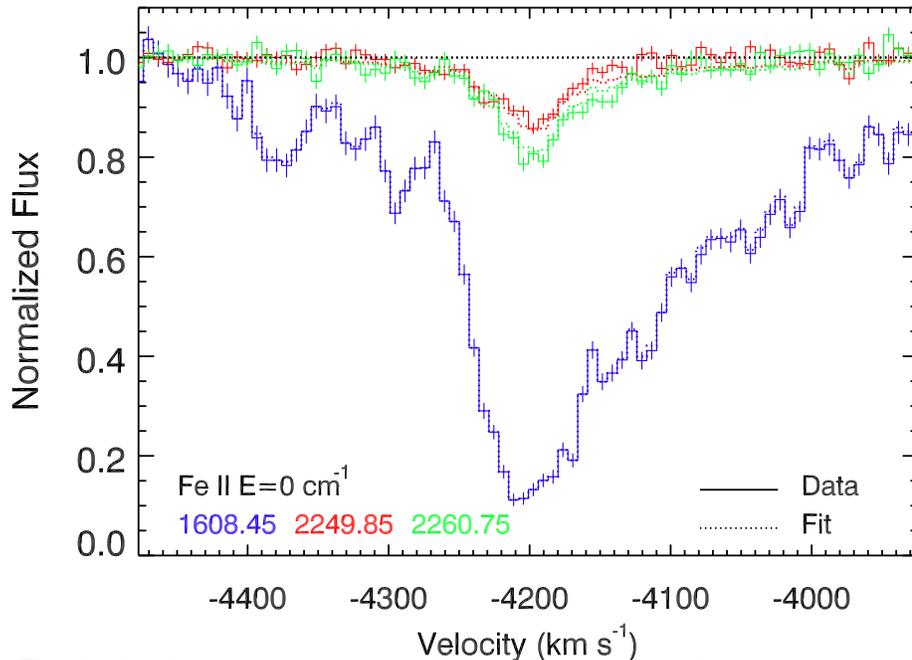}\\
  \caption[]
  {\footnotesize
The line fits for the three uncontaminated \feii\ resonance lines based
on the power law method. The data are presented in 7 km s$^{-1}$ bins
with a solid line (statistical observational error bars for these data
are also shown), while the model fit is shown by a dotted line.
Each line is color coordinated with the wavelength located at the lower
left of the plot.}
  \label{f3}
\end{figure}

We begin our column density determination by fitting the \feii\
resonance lines from the strongest component, ($\bf i$). In Figure 
\ref{f3}, we show the fits for three \feii\ resonance lines using 
the power law method. We select these particular line transitions 
because they provide us with a range of oscillator strengths and 
are free of blending or telluric contamination. Using these lines, 
we derive velocity dependent column densities that we integrate 
for the total ionic column density. 

Unfortunately, all strong \feii\ lines from the lower energy level 
385 cm$^{-1}$ are heavily blended in the UV 1, 2 or 3 multiplets with 
lines from both the resonance and other excited lines (e.g., \feii*
$\lambda$2396, shown in Figure \ref{f2}), which renders column 
density measurements based on these transitions to have large 
systematic errors. However, we find relatively strong lines outside 
of the UV 1, 2 or 3 multiplets for the 385 cm$^{-1}$ energy state
that are useful such as the \feii\ line $\lambda$1622 from the UV~38
multiplet. We also find two weaker lines from this energy state at 
$\lambda$1619 and $\lambda$2281 that are unblended. The oscillator 
strength values for these three lines are taken from Fuhr \& Wiese 
(2006). Of the three lines, both the $\lambda$1618 and $\lambda$1622 
are reported to have accurate oscillator strengths (rated as 
``A''\footnote{Accuracies of``A'', ``B'' and ``C'' are applied to
oscillator strengths with uncertainties less than 3\%, 10\% and 25\%,
respectively, according to the NIST Atomic Spectra Database} by Fuhr 
\& Wiese), while the weaker $\lambda$2281 has a 
reasonably reliable oscillator strength (rated ``B''). 
However, we find that using the oscillator strengths reported by Fuhr 
\& Wiese to be problematic. While the two weaker lines are well fit 
with these oscillator strengths, the trough depths relative to 
$\lambda$1622 are unphysical. For example, scaling the $\lambda$1619 
trough in apparent optical depth by the ratio of oscillator strengths 
($\sim$1.8) should be equal to the apparent optical depth in 
$\lambda$1622 (or larger in the case of partial covering), however 
the scaled optical depth of $\lambda$1619 is significantly smaller
than that of the $\lambda$1622 trough. This means that at least one 
of the oscillator strengths could be less reliable than previously 
believed. Because of this issue, we quote an ionic column density for 
the 385 cm$^{-1}$ energy level that is the mean of the fit to the 
$\lambda$1622 line alone and the fit to the two weaker lines with an 
uncertainty equal to the range in column density.

\begin{figure}[!h]
  \centering \includegraphics[angle=90,width=0.7\textwidth]
  {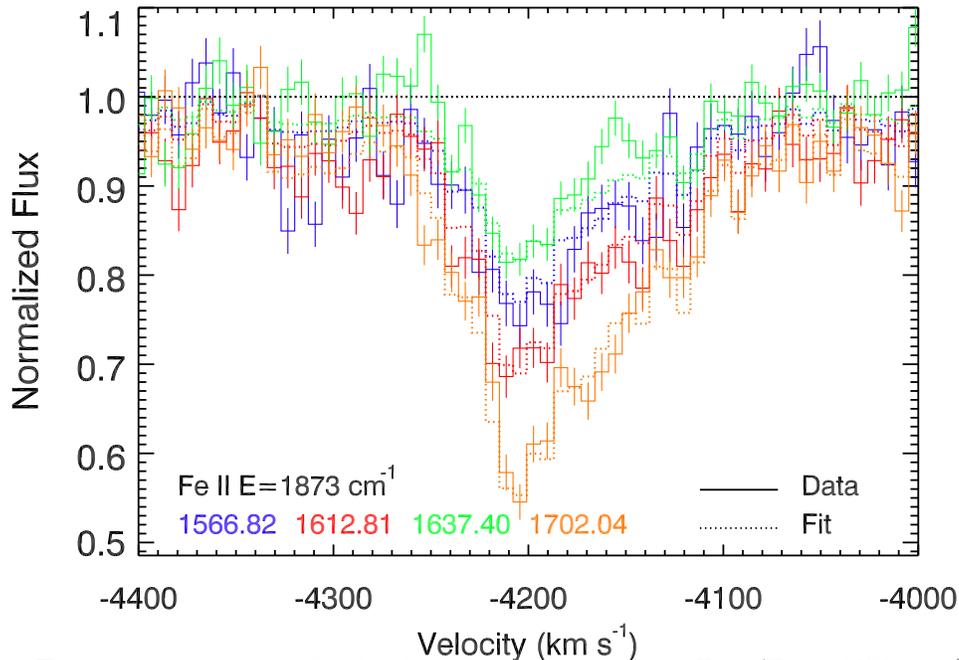}\\
  \caption[]
  {\footnotesize
The power law model fits for four lines of metastable
\feii\ (E$_{low}$=1873 cm$^{-1}$), we use the exponential $a$
parameter calculated from resonance \feii. Similar to Figure 3.}
  \label{f5}
\end{figure}

The other excited state lines from \feii\ (e.g., arising from 668, 
863, and 977 cm$^{-1}$) are also subject to heavy blending in the 
multiplets, but unlike the resonance and 385 excited state lines 
there is not sufficient column density to provide deep troughs with 
high signal-to-noise for the weaker lines. Thus, the column density 
measurements from these energy levels are significantly more 
uncertain. The exceptions to this are lines from the lower energy 
level state 1873~cm$^{-1}$ ($^4$F) and 7955~cm$^{-1}$ ($^4$D).

We find 5 separate absorption troughs of \feii\ arising from the 
level at 1873~cm$^{-1}$, which are shown in Figure \ref{f5}. We 
successfully fit these using the oscillator strengths from 
\citet{1995all..book.....K} to determine the column density for this 
level. However, the compilation of recommended oscillator strengths of 
\feii\ from Fuhr \& Wiese (2006) yields values lower by $\sim 0.3$~dex 
for two of the transitions, $\lambda$1637.4 and $\lambda$1702.0. For 
these transitions the recommended values are rated ``C'', while no 
oscillator strengths are given for the other two observed transitions. 
In addition to the difference in the absolute values, there is also a 
difference in the relative values, and in fact our fits to the absorption
troughs in the spectrum favors the relative values of Kurucz \& Bell (1995). 
Thus, we opt to use the relative oscillator strengths from Kurucz \& Bell 
(1995) to fit all four troughs, but the level column density derived this 
way is increased by 0.3~dex, as required by the absolute oscillator 
strengths of Fuhr \& Wiese. 

\begin{figure}[!h]
  \centering \includegraphics[angle=90,width=0.6\textwidth]
  {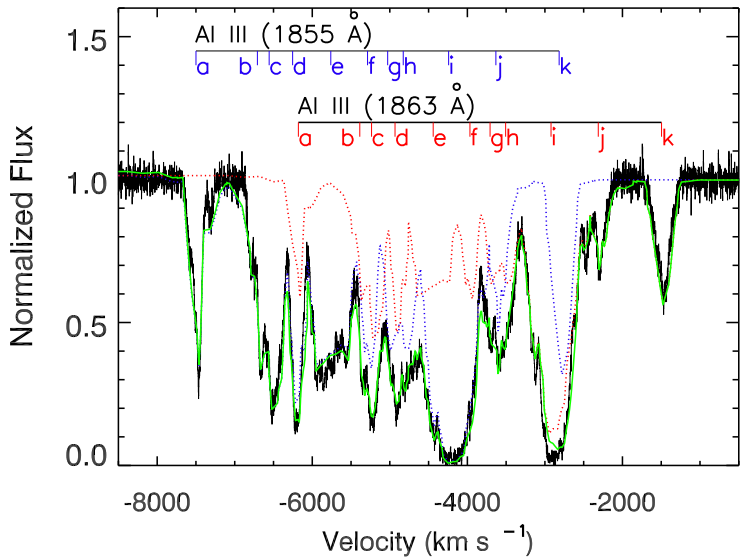}\\
  \caption[]
  {\footnotesize
The \aliii\ doublet plotted in velocity space with respect to the blue
member of the doublet. The blue and red
tickmarks and letters indicate the component positions of the \aliii\
doublet members, \aliii\ $\lambda$$\lambda$1855,1863 respectively.
The dotted blue and red lines are the individual \alii\ template
fits for each member of the doublet (corresponding to the blue and
red tickmarks). Each individual component is scaled in $\tau$ and the
resulting templates match the expected doublet ratio of 2:1 for each
of the components. We coadd, in $\tau$ space, the two \alii\ templates 
and present the resulting fit in green. The fit agrees with the data
with the exception of the $\lambda$1863 component $\bf i$, which is
likely due to saturation in the $\lambda$1855 counterpart.}
  \label{f8a}
\end{figure}

We find that the results from all three models (Power law, covering
factor, and AOD) are similar for the 
\feii\ column density measurements of component $\bf i$, which 
suggests that the LOS covering factor of the system is close to 1. 
Further evidence for a LOS covering factor$\approx$1 case is given 
in Figure \ref{f8a}, where we show that we can use the \alii\ template 
to fit the \aliii\ doublet. We scale two \alii\ templates
(one for \aliii\ $\lambda$1855 and one for \aliii\ $\lambda$1863) in 
$\tau$ space for each individual kinematic component for the blue and 
the red line of the doublet by the expected ratio of 2:1 and coadd the 
templates to generate a fit to the data. There exist only small 
deviations, and the overall fit to the doublet is good. The most 
obvious and important deviation is in the fit to the $\lambda$1863 
$\bf i$ component, which is due to some saturation in the core of the 
\alii\ template. This demonstrates that AOD is an adequate 
approximation and the LOS covering factor for all components is 
approximately 1.0. This same procedure gives similar results when 
applied to the \mgii\ doublet as well. 

\begin{figure}[!h]
  \centering \includegraphics[angle=0,height=0.6\textwidth,width=0.5\textwidth]
  {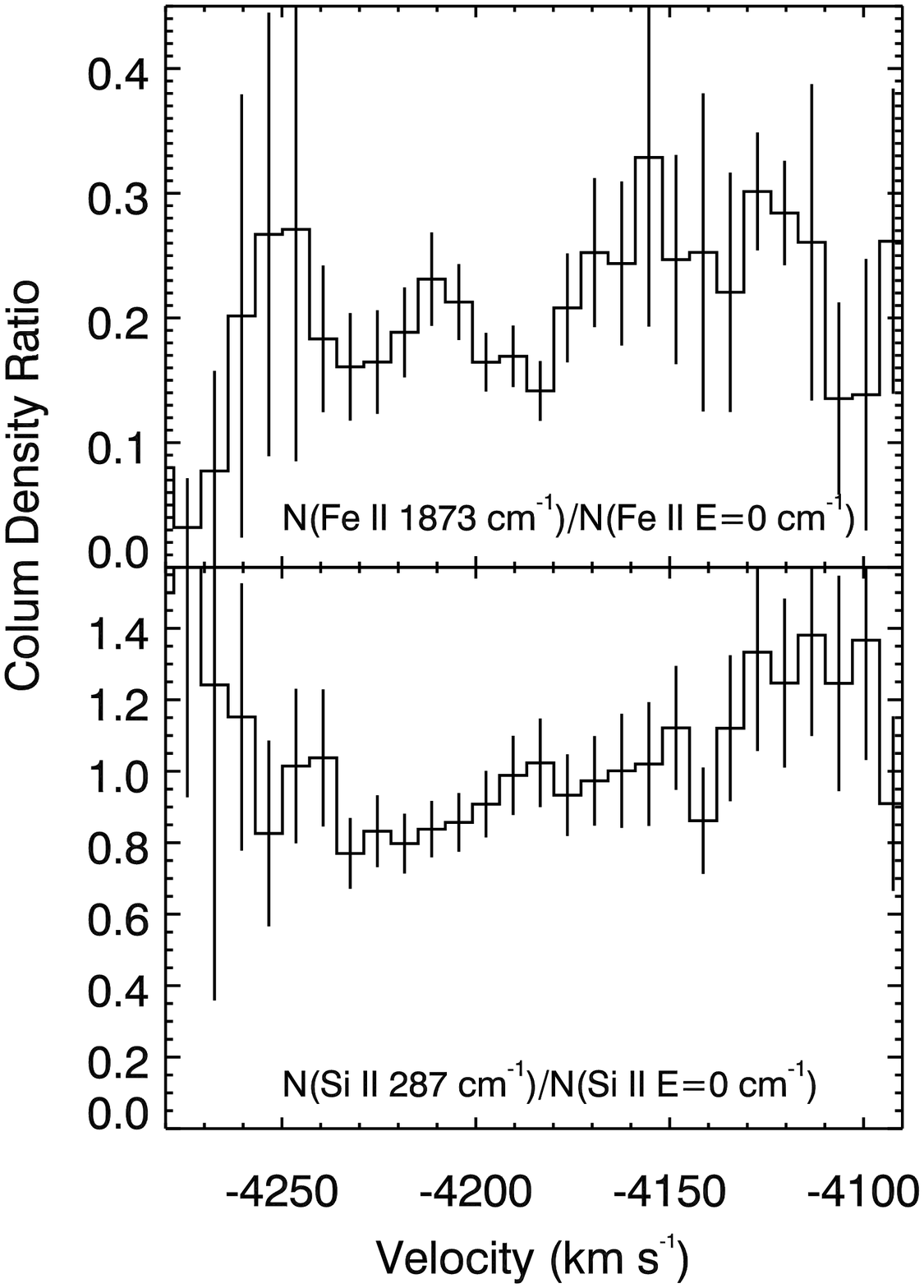}\\
  \caption[]
  {\footnotesize
Component $\bf i$ column density ratios of \feii*/\feii\ and 
\siII*/\siII, top and bottom respectively, in velocity space. 
Within the error bars the ratios are constant across the 
integration range. This indicates that the outflowing gas has 
a constant electron number density across the trough, which 
strongly suggests similar ionization conditions across the 
velocity span of this outflow component.}
  \label{f6}
\end{figure}

\subsubsection{The Physical Connection Between Material Seen in Different Troughs}

Ultimately, we want to establish the physical characteristics of
the outflow based on the column density determinations described
above.  In order to do so we use the integrated column densities as
input to photioinization models, which in turn determine the total
hydrogen column density and number density in the outflow.  This
procedure depends on two assumptions: 1) that the physical conditions
do not change as a function of velocity across a given trough. This
assumption allow for the use of integrated column densities instead of
working with each velocity resolution element separately (see Arav et
al 2007); 2) That the troughs of different ions correspond to the same
outflowing gas, which is a necessary condition for using these
measurements in the same photoionization models. Both assumptions can 
be checked using the data at hand.

The first assumption can be tested by checking the ratio measured
column densities amongst different lower levels of the same ion vary
as a function of velocity. In figure \ref{f6} we plot the ratios of
column densities between those arising from excited levels to those
from the ground state level of \feii\ and \siII. Within the
uncertainties, the ratios across the troughs are velocity independent,
showing that the electron number density is constant across the flow
(see \S~4 for elaboration).  This result gives validity to our first
assumption and justify the use of integrated column densities for
component {\bf i}. We also tested this for \niII/\niII* and
combinations between \niII, \siII\ and \feii\ resonance and excited
lines and find that the ratio is constant in all cases. This constant
ratio is also seen in the results for the outflow system in
QSO~2359$-$1241 as described by \citet{2008arXiv0807.0230K}.

\begin{figure}[!h]
  \centering \includegraphics[angle=0,width=0.7\textwidth]
  {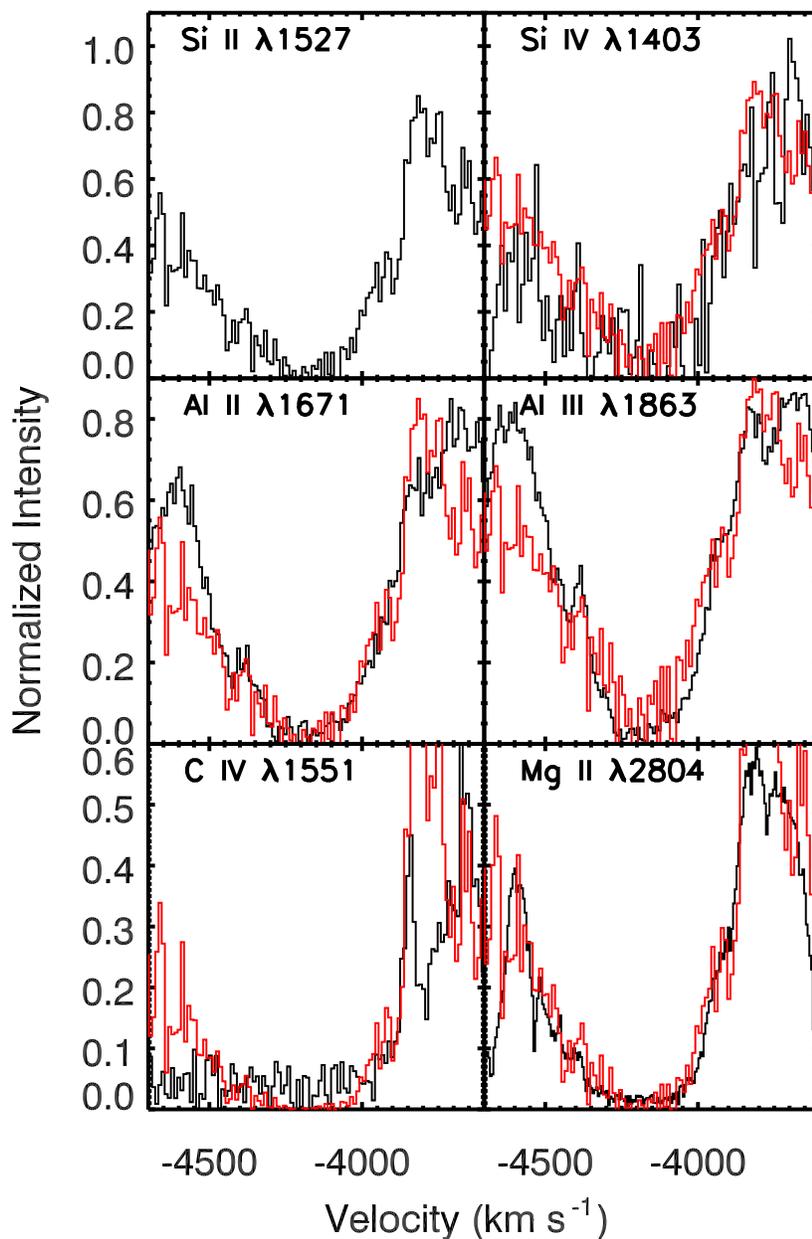}\\
  \caption[]
  {\footnotesize The top left panel shows the \siII\ $\lambda$1527 
$\bf i$ component. The remaining panels show the agreement of the 
\siII\ $\lambda$1527 $\bf i$ component trough with the red 
doublet member of 5 other ions. The \siII\ $\lambda$1527 data has
been converted to $\tau(v)$, where $\tau(v)$=-ln(I$(v)$), and
scaled by a single factor for each of the other lines in order to
produce the fits (red histograms) for these troughs (data shown in
black). Some discrepancy between the \siII\
$\lambda$1527 line and \aliii\ $\lambda$1863 occurs near 4000
km~s$^{-1}$, which is due to contamination from the blue $\bf k$
component. These agreements show that the ratio of column 
densities between each ion and \siII\ is constant across the 
profiles of the different troughs.}
  \label{f7a}
\end{figure}

In order to validate the second assumption we need to check the column
density ratio across troughs from different ions and especially
compare low and high ionization troughs. To do so we use the $\tau(v)$
profile of \siII\ $\lambda$1527 and attempt to fit the profiles of
other troughs while using a single scale factor for each line.  The
results are shown in figure \ref{f7a}. It is evident that the scaled
$\tau(v)$ profile of \siII\ $\lambda$1527 fits the troughs of \siiv,
\alii, \aliii\ troughs and \mgii\ remarkably well, lending strong
support for the assumption that we are measuring the same outflowing
gas.  For \civ, which is the highest ionization line observed, the fit
for the center of the trough is less conclusive since the trough is
highly saturated at its core.  However, we point out the the fit to
the low velocity wing (--3950 to -3850 \kms) is very good, which would
be a coincident unless the \civ\ trough can be fitted with the same
$\tau(v)$ profile of \siII\ $\lambda$1527.  In summary, the data
support the two main assumption that we need in order to use all
integrated column densities in our photoionization models.

\clearpage

\clearpage
\begin{figure}[!h]
  \centering \includegraphics[angle=90,width=0.9\textwidth]
  {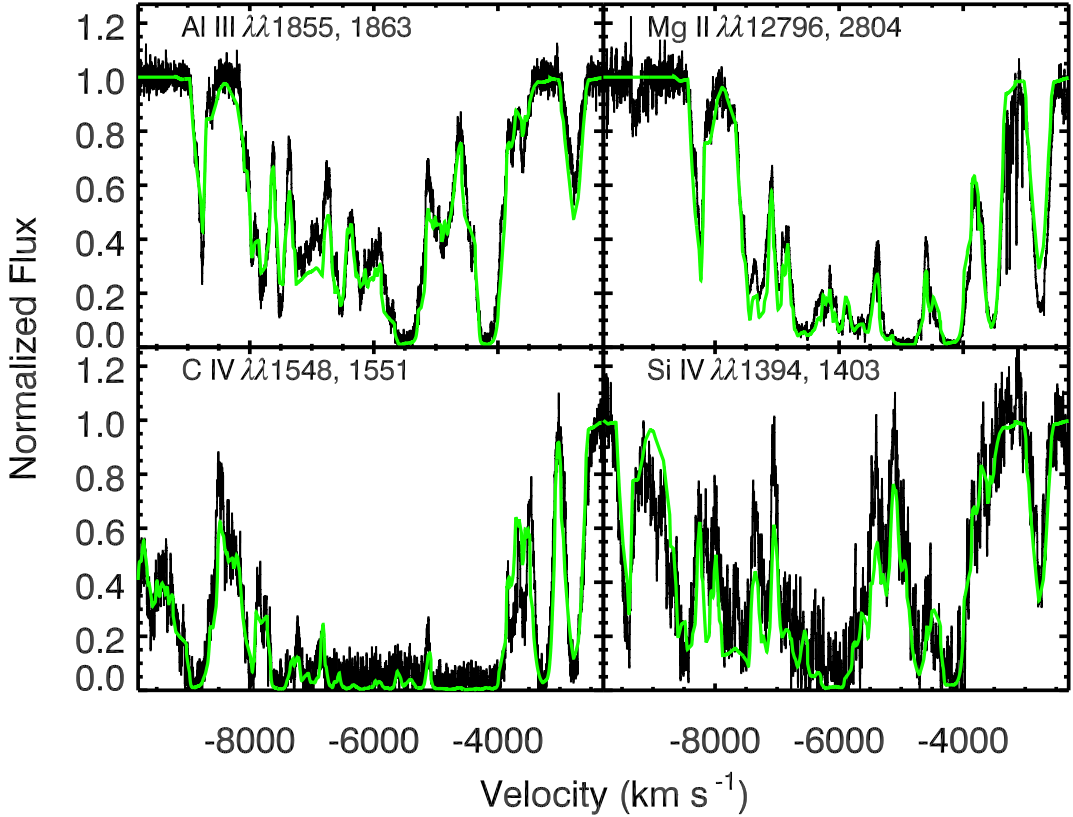}\\
  \caption[]
  {\footnotesize Fits to the \aliii, \mgii, \civ\ and \siiv\ doublets with
the composite \siII\ template for component $\bf i$ and the \alii\
template for all other components. We add two templates in a similar
fashion as for Figure \ref{f8a}. These fits, however, are scaled by
one constant for the blue template and scaled by 2:1 for the red
template. We also include the \siII\ $\lambda$1527 resonance line
and $\lambda$1533 excited line in the fit for \civ, which affect
velocities greater than $\sim-$5500. These 4 fits clearly demonstrate
that for each ion, all 11 components are present, and that no unexplained
components appear. We note that in the \civ\ fit there is evidence
for a component at $-$3600 km~s$^{-1}$ with a large optical depth
whereas the optical depth of this component \alii\ appears to be
significantly lower, thus the template fit is poor. The fits would
improve by scaling each component by a different factor in
optical depth.}
  \label{f7b}
\end{figure}

To demonstrate the tie between the high and low ionization troughs for
the other outflow components, we show in Figure \ref{f7b} that the 
combination of the \siII\ composite template (described in section 3.3 
for component $\bf i$) and the \alii\ template (for all other components) 
can be used to recreate the doublet troughs of \mgii, 
\aliii, \siiv\ and \civ. Similar to Figure \ref{f8a}, we add two 
templates together in $\tau(v)$. We scale the entire blue template (all 
11 components) by one factor and scale the respective red template by 
2:1 in $\tau(v)$. This method gives the fewest possible constraints for 
the fit. We find that all 11 kinematic components are present in both 
low and high ionization troughs, and that there are no unexplained 
kinematic components in the higher ionization state lines, which is
indication that this is the same outflowing gas for both the high and 
low ionization troughs. We also note that these fits could be better 
if we scaled each individual component for the blue member and kept 
the 2:1 ratio to the red, as we did for the \aliii\ fit in Figure 
\ref{f8a}. 

\subsubsection{Measurements of the Other Components}

The other two components with a sufficient number of measurable 
absorption lines are the $\bf k$ and $\bf a$ components. However, we 
are unable to utilize methods 1 or 2 due to a lack of unblended \feii\ 
resonance lines. For component $\bf k$, only one \feii\ resonance line 
(\feii\ $\lambda$1608) is measurable and unblended. Thus, we measure 
the column densities based on a LOS covering factor of 1. We further 
confirm this assumption of a LOS covering factor of 1, as with the 
$\bf i$ component, by checking that the $\lambda$1608 line scales well 
to the unblended portion of \feii\ $\lambda$2383 according to oscillator 
strength. Component $\bf a$ has no measurable \feii\ resonance lines. 
However, according to the \aliii\ doublet (Figure \ref{f8a}), we find that 
a LOS covering of $\sim$1 holds for all components and we are able to 
resort to the AOD method for our measurements. The results for components 
$\bf k$ and $\bf a$ are shown in Tables~\ref{compk} and Table~\ref{compa} 
respectively.

Due to the lack of measurable, unblended troughs for components $\bf b$ 
through $\bf h$, we use the apparent optical depth method, and 
integrate the total column density across the combined blended troughs 
for components $\bf b$ through $\bf h$, $-$7150 to $-$4600 km s$^{-1}$ 
in velocity, for the \siiv, \civ, \mgii, and \aliii\ doublets as well as 
the resonance and excited lines of \cii\ ($\lambda$1335, $\lambda$1336) 
and \siII\ ($\lambda$1527, $\lambda$1533). We present these results in 
Table~\ref{compbh}.

\begin{deluxetable}{lrcrrcc}
\tablecolumns{7}
%\tablewidth{1.2\textwidth}
\tabletypesize{\footnotesize}
\tablecaption{Component $\bf k$: Measured Column Densities}
\tablehead{
\colhead{Element} &
\colhead{Energy Level} &
\colhead{Velocity Range} &
\colhead{Column Density} &
\colhead{Uncertainty} &
\colhead{Number} &
\colhead{Methods$^a$}\\

\colhead{ } &
\colhead{(E$_{low}$ cm$^{-2}$)} &
\colhead{(km s$^{-1}$)} &
\colhead{($\times$10$^{12}$ cm$^{-2}$)} &
\colhead{($\times$10$^{12}$ cm$^{-2}$)} &
\colhead{of Lines$^b$} &
\colhead{ } \\
}

\startdata

\cii\   & 0    & $-$3088 $-$2249 & 1100     & 200  & 1 & 4 \\
\cii\   & 63   & $-$3088 $-$2249 & 1950     & 300  & 1 & 4 \\
\civ\   & 0    & $-$2893 $-$2641 & 1297     & 130  & 2 & 3 \\
\mgii\  & 0    & $-$2893 $-$2641 & 192      & 20   & 1 & 3 \\
\alii\  & 0    & $-$3047 $-$2494 & 35       & 4    & 1 & 3 \\
\aliii\ & 0    & $-$3047 $-$2494 & 73       & 7    & 1 & 3 \\
\siII\  & 0    & $-$3321 $-$2262 & 640      & 150  & 1 & 6 \\
\siII\  & 287  & $-$2893 $-$2641 & 352      & 40   & 1 & 3 \\
\siiv\  & 0    & $-$2893 $-$2641 & 140      & 20   & 1 & 3 \\
\feii\  & 0    & $-$2893 $-$2641 & 154      & 20   & 1 & 3 \\
\feii\  & 668  & $-$2893 $-$2641 & 9.0      & 1.0  & 1 & 3 \\
\niII\  & 0    & $-$3199 $-$2502 & $<$120   & ---  & 1 & 6 \\
\niII\  & 8394 & $-$2893 $-$2641 & 10.0     & 1.0  & 1 & 3 \\

\enddata

\tablenotetext{a}{Method used to determine column density, described
by the respective label number; 1: Power Law fitting, 2: Velocity
Dependent Covering Factor fitting, 3: Apparent optical depth fitting, 
4: ``wing'' fitting, 5: ``tip'' fitting, 6: Upper limit based on the 
noise method}

\tablenotetext{b}{Number of lines used in fitting. See Online Table
for specific transition information.}

\normalsize
\label{compk}
\end{deluxetable}

\thispagestyle{empty}

\begin{deluxetable}{lrcrrcc}
\tablecolumns{7}
%\tablewidth{1.2\textwidth}
\tabletypesize{\footnotesize}
\tablecaption{Component $\bf a$: Measured Column Densities}
\tablehead{
\colhead{Element} &
\colhead{Energy Level} &
\colhead{Velocity Range} &
\colhead{Column Density} &
\colhead{Uncertainty} &
\colhead{Number} &
\colhead{Methods$^a$}\\

\colhead{ } &
\colhead{(E$_{low}$ cm$^{-2}$)} &
\colhead{(km s$^{-1}$)} &
\colhead{($\times$10$^{12}$ cm$^{-2}$)} &
\colhead{($\times$10$^{12}$ cm$^{-2}$)} &
\colhead{of Lines$^b$} &
\colhead{ } \\
}

\startdata

\cii\   & 0    & $-$7523 $-$7271 & 333    & 30   & 1 & 3 \\
\cii\   & 63   & $-$7523 $-$7271 & 577    & 60   & 1 & 3 \\
\civ\   & 0    & $-$7558 $-$7243 & 734    & 70   & 1 & 3 \\
\mgii\  & 0    & $-$7523 $-$7271 & 28.7   & 2.9  & 1 & 3 \\
\alii\  & 0    & $-$7523 $-$7271 & 11.6   & 1.2  & 1 & 3 \\
\aliii\ & 0    & $-$7523 $-$7271 & 46.0   & 5.0  & 1 & 3 \\
\siII\  & 0    & $-$7523 $-$7271 & 101    & 10   & 1 & 3 \\
\siII\  & 287  & $-$7523 $-$7271 & $<$50$^c$ & ---  & 1 & 5 \\
\siiv\  & 0    & $-$7523 $-$7271 & 145    & 20   & 1 & 3 \\
\feii\  & 0    & $-$7695 $-$7300 & $<$40  & ---  & 1 & 6 \\

\enddata

\tablenotetext{a}{Method used to determine column density, described 
by the respective label number; 1: Power Law fitting, 2: Velocity
Dependent Covering Factor fitting, 3: Apparent optical depth fitting, 
4: ``wing'' fitting, 5: ``tip'' fitting, 6: Upper limit based on the 
noise method}

\tablenotetext{b}{Number of lines used in fitting. See Online Table
for specific transition information.}

\tablenotetext{c}{Column limit based on the limiting ratio ``tip'' fit 
for \siII*/\siII}

\normalsize
\label{compa}
\end{deluxetable}

\thispagestyle{empty}

%\newpage
\begin{deluxetable}{lrcrc}
\tablecolumns{5}
\footnotesize
\tablecaption{Components $\bf b-h$: Column Densities Limits}
\tablehead{
\colhead{Element} &
\colhead{Energy Level} &
\colhead{Velocity Range} & 
\colhead{Column Density} &
\colhead{Number} \\

\colhead{ } &
\colhead{(E$_{low}$ cm$^{-2}$)} &
\colhead{(km s$^{-1}$)} &
\colhead{($\times$10$^{12}$ cm$^{-2}$)} &
\colhead{of Lines$^a$} \\
}

\startdata

\cii\   & 0    & $-$6950 $-$4400 & $>$13000 & 1 \\
\civ\   & 0    & $-$7150 $-$4600 & $>$9900  & 1 \\
\mgii\  & 0    & $-$7150 $-$4600 & $>$910   & 1 \\
\alii\  & 0    & $-$7150 $-$4600 & $>$440   & 1 \\
\aliii\ & 0    & $-$7150 $-$4600 & $>$840   & 1 \\
\siII\  & 0    & $-$7150 $-$4600 & $>$3200  & 1 \\
\siiv\  & 0    & $-$7150 $-$4600 & $>$1900  & 1 \\
\feii\  & 0    & $-$7150 $-$4600 & $>$600   & 1 \\

\enddata

\tablenotetext{a}{Number of lines used in fitting. See Online Table
for specific transition information.}

\normalsize
\label{compbh}
\end{deluxetable}

\thispagestyle{empty}

\subsection{Estimates and Limits}

For situations where blending between components in doublets is 
prevalent (i.e. \civ, \siiv, \cii, and \cii*) or where the trough 
is heavily saturated (i.e., \civ), our primary method 
of column density extraction fails. Thus, we resort to other, less 
accurate, methods to estimate the column densities, 
which are important when constraining photoionization models. We 
develop three additional methods in order to estimate column 
densities in blended regions, which we explain here in order of 
decreasing accuracy.

\clearpage

In the first of our estimation methods (4), we opt to employ a 
template to map out approximate profiles for the troughs. We utilize
two templates for this process. The first template is a fit to the
\alii\ $\lambda$1671 line for all 11 components. However, due to 
moderate saturation in the core of the \alii\ $\bf i$ component, we 
create a composite template using the \siII\ resonance lines for 
the $\bf i$ component. 

To create this template, we use both the \siII\ $\lambda$1808 and
$\lambda$1527 lines. The $\lambda$1527 line is similar in shape to the
\alii\ $\lambda$1671 line, and while the core is moderately saturated,
it provides us with a good fit to the wings of the trough. In
contrast, the $\lambda$1808 line has a significantly weaker oscillator
strength and therefore the core of this trough is not saturated,
however the wings are buried in the noise. We therefore create a
composite template using the AOD column densities of the $\lambda$1527
for the wings (--3820 to --3990 \kms\ and --4390 to --4580 \kms), and
the AOD column densities of the $\lambda$1808 line for the core
(--4390 and -3990 \kms). We then convert the resultant \siII\ column
density profile to $\tau(v)$ template for the $\lambda$1527 line.

To measure column densities, we scale this composite $\tau(v)$
template to match the red wings of the \mgii, \siiv, and \civ\
doublets as well as \cii*, which are apparently free of blending. We
assume that the column density ratio is constant across the troughs,
which is supported by Figure \ref{f7a}, scale the composite template 
for the $\bf i$ component to match and then integrate the total column 
(assuming a
C$_{los}$$\approx$1) from the template. We refer to this method as the
``wing'' fitting method (method 4). The error estimation is provided
by using two $\tau$ scaling values, one underestimating and one
overestimating the noise level in the trough's wing.  We note that
these measurements use only one line, the red doublet member, which is
reflected in Tables~\ref{compi} through \ref{compbh}. To further test
the validity of this method, we used the composite template to fit the
wing of the \siII\ $\lambda$1527 line and we find that the template is
able to reproduce the column density we find in our power law fitting
of the $\lambda$1808 line to within 10\%. 

The next method we designate as the ``tip'' fitting method, or method 
5, where we estimate the column densities in a similar 
way to method 4. For example, we estimate the column density of \oi\ 
$\lambda$1302 for the $\bf i$ component in the region containing 
blended lines from \siII\ $\lambda$1304 and \siII* $\lambda$1309. We 
find that the composite template fit for the $\bf i$ component agrees 
well with the deep trough at $\sim$3809 \AA\ in velocity space (shown
in online figure 1), which 
appears to be the tip of \oi\ absorption. Thus, we use the composite 
\siII\ template to approximate the minimum and maximum amount of column 
density that could come from \oi. Due to the uncertainty in continuum 
placement and possible saturation, we consider this method less accurate 
than method 4, which is reflected in the larger error bars.

Our last column density estimation method (the noise method or method 
6) defines upper limits on the column density of ions based on the 
statistical error of the data. Because the column density of \hei\ is 
important for photoionization models \citep[see][]
{2001ApJ...546..140A,2008arXiv0807.0230K}, we estimate an upper limit 
for the column density of the excited 2$^3$s level of \hei\ (\hei* 
hereafter) in component $\bf i$ for \hei* $\lambda$3189. To do this, 
we scale the \siII\ template for the $\bf i$ component and compare it 
to the region where the line would be detected. We vary the scaling 
to the noise level, find the largest possible optical depth that could 
be buried without detection and integrate across the template to determine 
the column density. We also perform this analysis for \mgi\ $\lambda$2853, 
\fei\ $\lambda$2484, and \ci\ $\lambda$1560; and quote the measured limits 
in Table~\ref{compi}. 

Component $\bf k$ presents its own problems. Unlike the $\bf i$ 
component, here we effectively measure column densities for \feii\ 
resonance and excited levels (E=0 cm$^{-1}$ to E=668 cm$^{-1}$), but 
we lack a \siII\ resonance line to provide a ratio to the \siII\ 
$\lambda$1309 excited line. Therefore, we apply method 5 with the 
\alii\ template to find limits on the \siII. Moreover, we use the 
``wing'' fitting method (method 4) with the same \alii\ template to 
estimate the column density for \cii\ and \cii* in component $\bf k$.

In component $\bf a$, the number of measurable lines is even fewer 
than for component $\bf k$. The only clean, unblended density 
diagnostic is \cii\ to \cii*. As the critical density for this ratio 
is low \citep[$\sim$10$^2$ cm$^{-3}$,][]{1992ApJS...80..425B,
1983JPHYSB..16..157},  it is important to find another density check. 
Thus, using the \alii\ template, we estimate an upper limit to 
the column density ratio of \siII* to \siII\ to be 0.49 for component 
$\bf a$. 

\subsection{Systematic Uncertainty}

We recognize 
that there exist systematic errors that could contribute further to
the uncertainty in our measurements. For example, our placement of the
continuum, defined in Section 2, could result in increased errors of
the \siII\ $\lambda$1808 and \siII* $\lambda$1817 column densities by
5\%. The amount of saturation in the core of \siiv\ is another
systematic error that could be a potential source of error in column
density, which at maximum, could lead to an error in ionization
parameter of 0.1 dex. Both of these are negligible compared to the
uncertainties due to the chemical abundance choice.

\section{Spectral diagnostics and modeling} 

The observed spectrum of SDSS~J0318$-$0600 exhibits distinct absorption 
from \cii\ and \civ, \oi, \mgii, \alii\ and \aliii, \siII\ and \siiv, 
\crii, \feii, \coii, \niII, and \znii\ for component $\bf i$. In 
addition, we measure upper limits to the column densities of \hei*, 
\fei, \ci\ and \mgi. 
As already shown, a similar absorption trough structure is shared by 
the strongest transitions of the singly ionized species, and those of 
intermediate ionization (Al III and Si IV), as well as C IV. 
Additionally, component $\bf i$ likely has the largest column density 
of any component in SDSS~J0318$-$0600, given that it has the largest 
number of absorption line identifications and is the deepest absorption 
feature for transitions of low oscillator strengths and/or low relative 
elemental abundance. Therefore, we analyze component $\bf i$
here and analysis of all other absorption components will be 
presented in a subsequent paper (Bautista et al. 2009, in prep).

One of the main objectives of modeling the observed ionic column
densities is the determination of the total hydrogen column density  
of the outflow $N_H$ and its distance from the central source $R$, 
which are key in the determination of kinetic luminosity of the 
outflow (see Section 6). The former is obtained directly through 
photoionization modeling, but the latter is not. Instead, modeling 
yields the so called ionization parameter defined as:
\begin{equation}
U_H = {Q_H\over {4\pi R^2 \vy{n}{H} c}} = {\Phi_H\over{\vy{n}{H} c}}
\end{equation} 
where  $\vy{n}{H}$ is the total hydrogen density, $c$ is the
speed of light, $\Phi_H$ is the surface flux of 
ionizing photons, $Q_H$ is the rate of hydrogen ionizing 
photons emitted by the central object, which is defined by: 
\begin{equation}
Q_H = \int_{1 Ry}^{\infty} \frac{L_\nu}{h\nu} d\nu~{\rm (ionizing~photons~s^{-1})},
\end{equation}
where the integration range covers the range of ionizing photons,
L$_{\nu}$ is the luminosity at each frequency and $h$ is the Planck 
constant. From this definition, $Q_H$ can be estimated directly from 
the observed luminosity ($\lambda$L$_{\lambda}$ of the object and some 
knowledge of its spectral energy distribution (SED), see Sections 4.2 \& 
4.3). Thus, the distance $R$ can be obtained from the diagnosed value of 
$U_H$ whenever one has an independent determination of $\vy{n}{H}$, which 
in a cloud with fully ionized hydrogen is roughly equal to the electron 
density, $n_e$.

Thus, the first step in the analysis is to determine $n_e$ from its 
effect via collisions on the relative populations of excited levels 
observed in the spectrum. Later, we discuss the nature of the SED, which 
is to be used in the determination of $Q$ and as input for photoionization 
modeling. After this, we will proceed to detailed photoionization modeling.

\subsection{Electron density determination}

Under excitation equilibrium conditions the column densities of lowly excited
metastable levels can be used as diagnostics of electron density,
assuming the atomic parameters and excitation processes responsible
for excitation are sufficiently well understood. We have measured column densities
for the 0 cm$^{-1}$ ground state and low-lying metastable excited levels within
\siII, \feii, and \niII.

\begin{figure}[!h]
\centering \includegraphics[width=10.0cm,height=11.0cm,angle=0]
{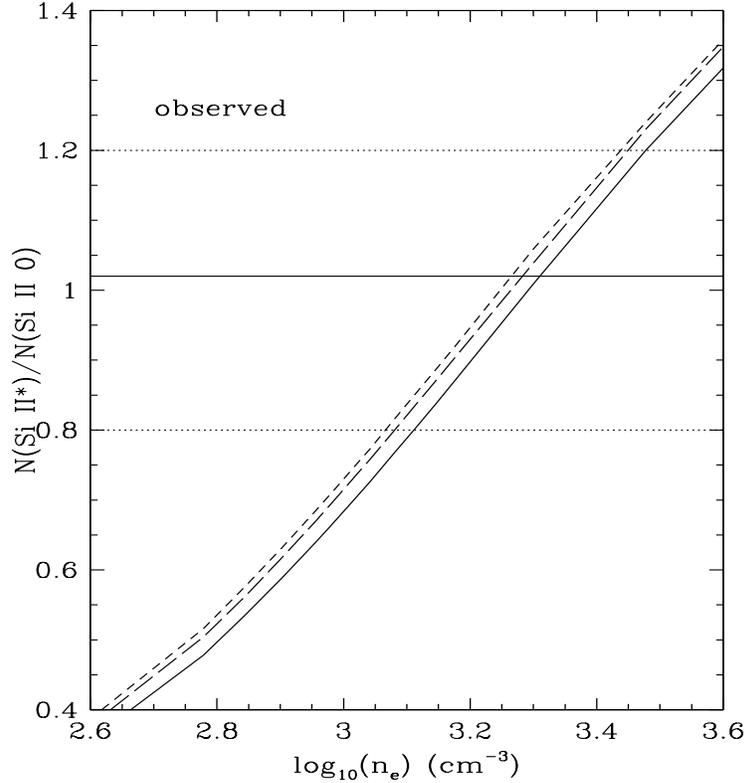}\\
\caption{\footnotesize
Electron density diagnostic ratio of the populations of the
first excited level of \siII\ and the ground state. The theoretical line ratios
are calculated at $T_e=10,000$K, although the ratio is rather insensitive to this
choice. The various curves depict the ratios obtained from the
collision strengths of \citet[][solid curve]{1991MNRAS.248..827D},
\citet[][long-dashed curve]{2008ApJS..179..534T}, and Bautista et al.
(2009, in prep., short-dashed curve). The horizontal lines indicate the
observed ratio (heavy solid) and estimated uncertainties (dotted), which
are dominated by the uncertainties in the oscillator strengths of the 
transition pair at 1808.0 \AA\ and 1816.9 \AA.}
\label{si2fig}
\end{figure}

The simplest and probably most reliable of the electron density diagnostics
is that from the ratio of the column densities of the \siII\ levels, as shown in
Figure~\ref{si2fig}. Here, the ratio of the excited 3s$^2$3p~$^2$P$^o_{3/2}$
at 287 cm$^{-1}$ to the 3s$^2$3p~$^2$P$^o_{1/2}$ ground level is
calculated from three different sets of effective collision strengths,
\citet{1991MNRAS.248..827D}, \citet{2008ApJS..179..534T}, and Bautista
et al. (2009, in prep.). We measure this column density ratio from the pair of
absorption lines at 1808.0 \AA\ (resonant) and 1816.9 \AA\ (excited state). As
described above, we have a good deal of confidence in their measured integrated
optical depths, especially their ratio, and we will thus place the greatest weight
on this electron density diagnostic. We caution, however, that as the weakest amongst
such pairs of transitions commonly
measured in \siII, there are moderate uncertainties in their oscillator
strengths, even in their relative values (about 0.1 dex). Two other such pairs of
transitions that lie within our spectrum, centered on 1307 \AA\ and 1531 \AA,
have smaller uncertainties in their oscillator strengths but are saturated 
in our spectrum and are not useful for direct measurements. These uncertainties 
dominate over the statistical
measurement errors in the column density fits to the observed troughs at 1808.0 \AA\
and 1816.9 \AA.~Figure~\ref{si2fig} illustrates the electron density dependence
of this level specific column density ratio, wherein the solid and dotted horizontal
lines represent our measurement of this ratio and an estimated range of its
uncertainty, the latter dominated by the uncertainties in the transition oscillator
strengths. We derive an electron density of $\log(n_e) = 3.3\pm0.2$ (cm$^{-3}$).

\begin{figure}[!h]
\centering
\includegraphics[clip=true,viewport=.0in .0in 4.5in 12in,angle=-90,width=7.5in]  {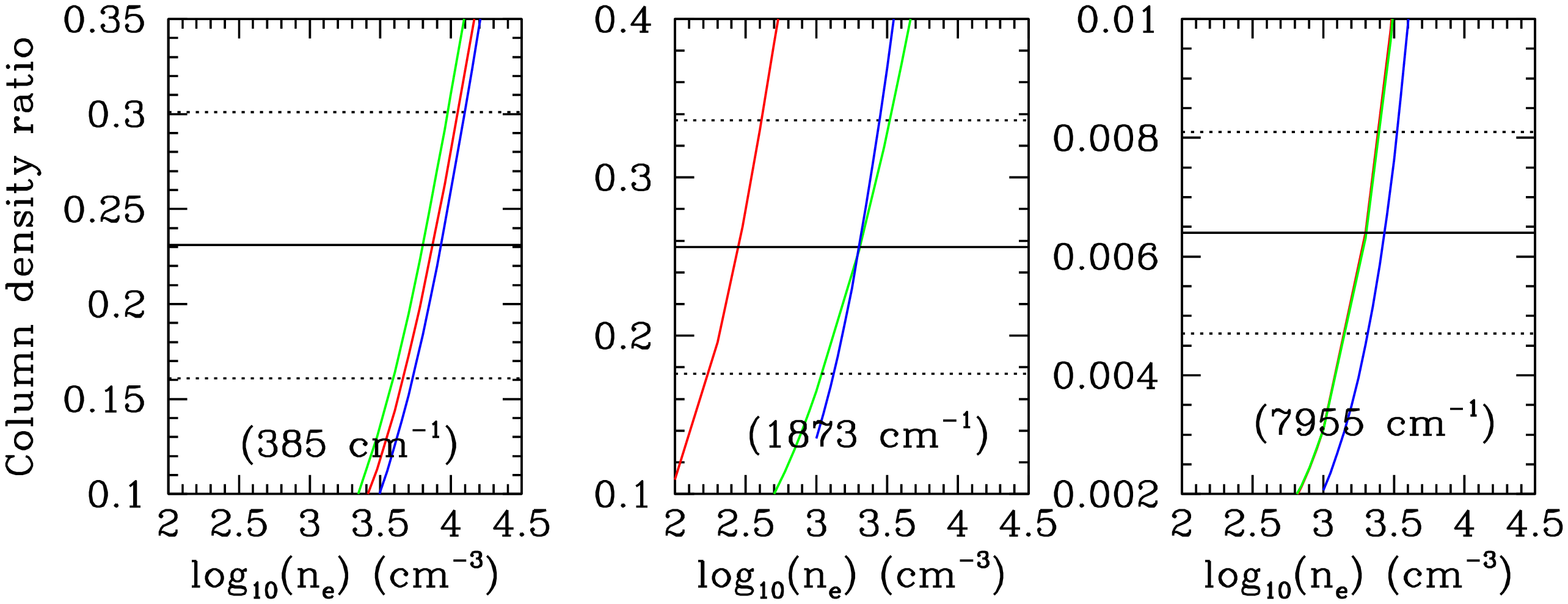}\\
\caption
{\footnotesize Electron density diagnostics from the relative populations
of excited metastable levels of \feii. The ratios are: $N$(\feii* 385)/$N$(\feii\ 0),
$N$(\feii* 1873)/$N$(\feii\ 0), and $N$(\feii* 7955)/$N$(\feii\ 0).
The various curves depict the predicted
line ratios by pure collisional excitation according to
\citet[][red]{1998ApJ...492..650B}, by full {\sc Cloudy} photoionization models 
dominated by collisional excitation 
\citet[][blue]{1999ApJS..120..101V}, and 
a model that accounts for \civ\ Bowen fluorescence (green). The
measured column density ratios are depicted by horizontal solid lines,
while the corresponding errors are indicated by dashed lines. We note that
the red curve for the $N$(\feii* 7955)/$N$(\feii\ 0) ratio is located 
underneath the green line.}
\label{fe2fig}
\end{figure}

The various observed transitions arising from low-lying metastable levels of the \feii\ 
ion allow for independent electron density diagnostics. However, constraining the 
electron density from the relative populations within these metastable levels 
proved more difficult. We present these in 
Fig.~\ref{fe2fig} for column densities of the 385~cm$^{-1}$, 
1873~cm$^{-1}$, and 7955~cm$^{-1}$ levels. These are the only levels for which
we have sufficient confidence in their measured column densities (see Table 2) 
to justify their utility as electron density diagnostics. In blue are the 
predictions from {\sc cloudy} \citep[v06.02,][]{1998PASP..110..761F} for photoionization 
models presented in Korista et~al. (2008) for the main outflow system in QSO~J2359$-$1241 
(solar abundances, fixed ionization parameter and total column density for which 
no hydrogen ionization front formed, and $n_e \approx \vy{n}{H}$ within the Fe$^+$ zone). 
Relevant mainly for the ratios involving the 7955~cm$^{-1}$ level, the electron 
density weighted average temperature within the Fe$^+$ zone of these models was 
approximately 
11,000~K. In red are the predictions of the collisional equilibrium model \feii\ 
atom of \citet{1999ApJS..120..101V} for a fixed electron temperature (10,000~K) at 
each electron density. This difference in adopted temperature has no
significant effect on the populations of the 385~cm$^{-1}$ and 
1873~cm$^{-1}$ levels, while the population of the 7955~cm$^{-1}$ level is
roughly linearly dependent on electron temperature for the conditions of interest 
(e.g., $\sim 10\%$ differences are found for the two temperatures quoted). 
The horizontal solid and dashed lines are the measured 
values and estimated uncertainties, respectively. The small differences in the 
predictions pertaining to the 385~cm$^{-1}$ level and the large differences pertaining 
to the 1873~cm$^{-1}$ level are primarily due to differences in the radiative
transition rates (A-values) for forbidden transitions used in the two sets of 
computations. A detailed inspection of the atomic 
data files of the \citet{1999ApJS..120..101V} model atom in {\sc cloudy} revealed 
that for a few dipole forbidden transitions the model adopts A-values whose sources 
are neither known to us nor cited in their publication. For all other relavent 
transitions the Verner et~al. model atom in {\sc cloudy} adopts the A-values 
from \citet{1996A&AS..120..361Q} \footnote{This situation is present within 
{\sc cloudy} versions 08 and prior, while future versions are expected to adopt a 
revised \feii\ model (Ferland 2009, private communication)}. 
The largest effects of the A-values of unknown origin present in {\sc cloudy} are 
with respect to the population of the 1873~cm$^{-1}$ level, where the differences 
are substantial as indicated in the middle panel of Fig.~\ref{fe2fig} (blue and red 
curves). \footnote{We have verified that the effect of the present discrepancy in
the atomic data for \feii\ on the derived gas density in the main outflow system of 
SDSS~J2359$-$1241 reported in Korista et al. (2008) is minimal}. 

The electron density predicted by the \siII\ lines ($\log(n_e) = 3.3\pm0.2$ cm$^{-3}$, 
see Figure~\ref{si2fig}) and that from the 385~cm$^{-1}$ level in \feii\ 
($\log(n_e) \approx 3.9$ cm$^{-3}$) differ somewhat in value. However, 
considering the moderate uncertainties in the relevant atomic data of the \siII\ and 
\feii\ 385~cm$^{-1}$ transitions available to us as diagnostics of electron 
density, and the larger uncertainties in the measurements in the \feii\ 385~cm$^{-1}$ 
troughs, we do not consider these differences to be significant.

In contrast, the predicted electron density based on the relative population of the
\feii\ a~$^4$F$_{9/2}$ level at 1873~cm$^{-1}$, being an order of magnitude smaller,
does present a possible conundrum. Among all of the column density measurements of 
transitions from excited states of \feii\, we believe this one to be the most secure. 
However, the atomic data (A-values and 
collision strengths) involved in the computations of level populations
of the a~$^4$F term carry
greater uncertainties than those 
pertaining to the levels of the a~$^6$D (ground) term. 
This is because the a~$^4$F and a~$^6$D terms belong to different electron
configurations, i.e. 3d$^7$ and 3d$^6$4s respectively, and the energy difference
between these terms is comparable in magnitude to the relativistic corrections to the
Hamiltonian. Thus, the computation of accurate transition rates among levels of these
terms requires fine tuning of large atomic configuration expansions (see
Bautista 2008 for a more detailed discussion).

We also consider the possibility of some selective excitation process that might 
lead to a significantly lower population in \feii\ a~$^4$F$_{9/2}$ than would be
inferred by collisions alone. Neither 
photoexcitation by the continuum radiation field nor Ly$\alpha$ fluorescence 
can explain the observed discrepancy. Another mechanism, known to occur 
in spectra of circumstellar envelopes of symbiotic stars and in the environs of
gamma-ray bursts is Bowen fluorescence 
of \feii\ by \civ\ emission at 1548 \AA\ \citep[e.g.][]{1983MNRAS.205P..71J, 
2008A&A...477..255E,2007A&A...468...83V,2000A&A...359..627H,1992ApJ...389..649M}. 
Pumping occurs as the 1548.19 \AA\ resonance line 
of \civ\ closely coincides with the 1548.20 \AA\ transition from the 
a~$^4$F$_{9/2}$ metastable level to the y~$^4$H$_{11/2}^o$ level of 
\feii.~Subsequently, this high level cascades down to the ground stage 
of the ion, so the population of the a~$^4$F$_{9/2}$ level is 
re-distributed among other low lying metastable levels 
\citep[see also][]{2000ApJ...543..831V}. 
Fully self-consistent modeling of this fluorescence process requires a 
detailed treatment of radiative transfer of \civ\ and \feii\ lines 
within the cloud, which is beyond the scope of the present paper. 
Nevertheless, we can estimate the relative level populations of the observed levels 
in our spectra under the fluorescence scenario by artificially including its 
effect in the Bautista \& Pradhan (1998) collisional equilibrium \feii\ model. 
We find that solving this discrepancy 
requires a photoexcitation rate in the 1548.20 \AA\ transition that is at least 
ten times greater than what is available through photoabsorption of radiation 
from the adopted dereddened SED (Section 4.2). We show
the resulting effects on the level populations as the green curves in 
Fig.~\ref{fe2fig} (i.e., the red curve becomes the green one).
Although, this
fluorescence mechanism is able to correct the population of the
a~$^4$F$_{9/2}$ level to agree with the observed column density at
the same electron density given by \siII, the pumping rate required
seems too large even 
when considering the inclusion of \civ\ emission from the broad emission line region.
Further investigation of this effect is left for future work. 
After considering all of these issues pertaining to the various uncertainties,
we adopt the \siII-determined electron density of $\log(n_e) = 3.3\pm0.2$ (cm$^{-3}$).

Finally, we measure column densities from the ground 3d$^9$ $^2$D$_{5/2}$ level
and the 3d$^8$4s $^4$F$_{9/2}$ metastable level at 8394 cm$^{-1}$ for the \niII\ ion.
We find that the population of the metastable level exceeds the predictions
of pure collisional excitation conditions for any value of the electron
density. This is to be expected because \niII\ is very efficiently excited
by continuum radiation \citep{1996ApJ...460..372B,1995A&A...294..555L}. As  
a consequence \niII\ cannot be used as density diagnostic here, but 
instead serves as a check for the flux of photons below the hydrogen
ionization threshold from the central source. As such, \niII\ can be  
used as a direct distance indicator to the outflow (see Section 4.4).

\subsection{The Spectral Energy Distribution}

The SED incident on the outflowing
gas has important consequences for the ionization and thermal structures
within the gas. In what follows, we provide motivation for our choice of 
two standard SEDs to use in the photoionization models. While AGN are
routinely observed in the optical/UV and X-rays
\citep[see][]{1994ApJS...95....1E}, the critical EUV spectral region 
($\sim$1~Ry to 0.1~keV, rest frame) is largely inaccessible except near 
its low and high energy extremes. The present working model is that a
thermal accretion disk spectrum dominates the energy contained within
the continuum spectrum feature known as the ``UV bump'', sometimes
called the ``big blue bump'' \citep{1978Natur.272..706S,1983ApJ...268..582M}. 
Such a feature is expected to peak (in units of $\nu F_{\nu}$) at energies 
of a few Rydberg, depending on the mass of the supermassive black hole and
the mass accretion rate \citep{1973A&A....24..337S}. 
\citet{1987ApJ...323..456M} had this in mind when they constructed a 
simplified piece-wise
power-law representation of a typical QSO SED, basing the relative
strength and roll-over energy of the UV-FUV bump on measurements of the
strengths of the \heii\ broad emission lines (photon counting) and overall
energy balance within the broad emission line gas (e.g., comparing the
strength of Ly$\alpha$~1216 to the strengths of strong cooling lines
such as \civ\ $\lambda$1549), and joining the turn-down onto a power-law
representation of the X-ray spectrum for typical X-ray bright quasars
(samples of which at that time were dominated by radio loud QSOs). This
spectrum, hereafter referred to as MF87, has since been the standard SED
in modeling photoionized gas associated with AGN. We adopt the MF87 SED
as one of two standards for our photoionization models.

\begin{figure}[!h]
%\rotatebox{-90}{\resizebox{\hsize}{\hsize}
%{\plotone{figures/contfig.eps}}}
\centering \includegraphics[angle=-90,width=0.8\textwidth]
{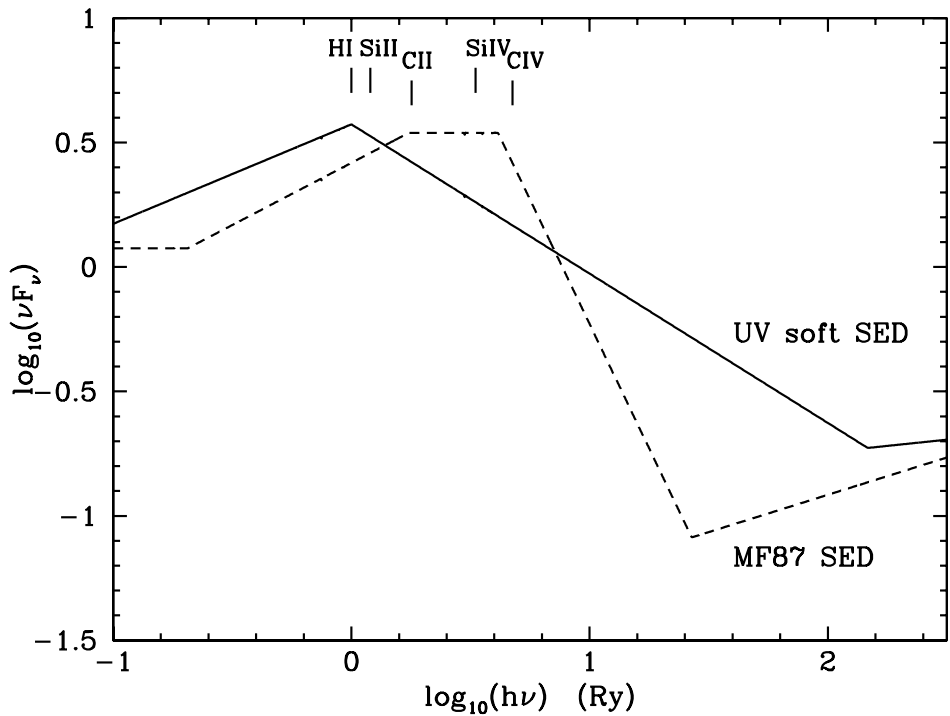}\\
\caption{\footnotesize The incident SEDs considered in photoionization 
modeling. Both SEDs shown have the same bolometric flux. 
Here we also mark the ionization energies of some ions
of interest (i.e. \hi, \siII, \siiv, \cii, \civ).} 
\label{contfig}
\end{figure}

By the late 1990s, improved UV and X-ray observations of AGN led several
investigators to reconsider the importance of accretion disk emission
to the FUV spectra of AGN. In a sample of luminous QSOs with {\em
HST} spectra, \citet{1997ApJ...475..469Z} found surprisingly soft continuum 
SEDs at wavelengths shortward of $\sim$1000 \AA. Shortly thereafter,
\citet{1997ApJ...477...93L} reported that the soft X-ray spectra of quasars
appeared to be pointing (in $\nu F_{\nu}$ space) at the soft FUV slopes
found by Zheng et~al.\, and proposed that a large fraction of the UV bump
inferred by \citet{1987ApJ...323..456M} and standard accretion disk models
was missing \citep[see also][]{1994ApJS...95....1E}. This led to 
investigations of cooler accretion disk models with thermal spectra that 
are Comptonized
by a $\sim 10^8$~K plasma, which result in power-law like instead of
exponential like high energy tails in the spectrum 
\citep[e.g.][]{2001ApJ...563..560B}. \citet{2002ApJ...565..773T} 
followed up on the work of Zheng et al.\
with a larger sample of objects with {\em HST} spectra, which improved
the statistics at the shorter wavelengths. They found similarly but
not quite as soft FUV slopes, and concluded for radio quiet quasars
``that it is plausible to represent the entire typical ionizing
continuum from $\sim 10$~eV to $\sim 2$~keV by a single power law''
($\alpha \approx -1.6$). With this in mind \citet{2001ApJ...550..142H,
2002ApJ...564..592H} constructed a simple piece-wise power law radio quiet
quasar SED to model intrinsic AGN absorption systems and broad emission
lines. We adopt a slightly modified version of their SED as our
second standard: from 1 micron to 1~Ry, $\alpha = -0.6$; from 1~Ry
to 2~keV, $\alpha = -1.6$; from 2~keV to 50~keV, $\alpha = -0.9$ and
a high energy break beyond. This SED has an $\alpha_{ox} \approx -1.43$,
very nearly equal to that of the MF87 SED ($\alpha_{ox} \approx -1.40$),
while effectively excluding the presence of a UV bump peaking at FUV
energies\footnote{Hamann et al. 2001 terminated this power law 
portion at 1 keV, resulting in a slightly flatter alpha$_{ox}$ = -1.35.}. 
We refer to this as the "UV-soft SED", and we show it along with the 
MF87 SED in Figure~\ref{contfig}.

%\clearpage
%\clearpage

With regards to these observational studies of quasar SEDs, the reader
should bear in mind that significant uncertainties remain in the
corrections in the UV for Galactic extinction (primarily in the adopted 
value of
$R_V$), in the corrections for atomic absorption within the intervening
IGM (mainly the Ly$\alpha$ forest), as well as in the quasar host galaxy
extinction. No attempt was made to correct for the last of these in
any of the above works. Thus from an observational point of view, the
existence and magnitude of the 1000~\AA\/ break and the apparent roll-down
toward the observed X-rays, as suggested in \citet{1997ApJ...477...93L}, 
remains unclear. \citet{2004ApJ...615..135S}, in a sample of {\em FUSE} 
spectra of luminous Seyfert1 galaxies and low redshift quasars, reported 
{\em no} FUV spectral break down to $\sim$600~\AA\/. However, some of the 
differences between this finding and that in the more
luminous AGN sample of Telfer et~al. (2002) may
be intrinsic, due to differences in the accretion disk physics. We refer 
the reader to \citet{2005ApJ...619...41S} and \citet{2007ApJ...659..211B} 
for recent detailed comparisons between observations and modern accretion 
disk model predictions, as well as critical evaluations of the current 
uncertainties. We consider it plausible that actual quasar FUV/EUV SEDs 
fall somewhere between our two adopted standards.

\subsection{Reddening of the SED}

In the previous section, we constructed a plausible UV-soft SED for 
SDSS~J0318$-$0600.
However, it is known that the UV spectrum of SDSS~J0318$-$0600 is strongly 
affected by extinction due to dust. \citet{2002ApJS..141..267H}
found that if the Small Magellanic Cloud reddening curve, typical for QSOs, 
were adopted the extinction correction $E(B-V)$ would be between $\sim$0.1 
and $\sim$0.4 depending on whether the 2000-3000 \AA\ or 1250-2000 \AA\ rest 
frame wavelength ranges were fit to the SDSS composite QSO spectrum. 
Because the location of the bulk of the reddening of the SED is unknown, 
we must consider a scenario in which the SED incident on the outflow has
been reddened by dust grains. 

\begin{figure}[!h]
%\rotatebox{0}{\resizebox{\hsize}{\hsize}
%{\plotone{redningfig.eps}}}
\centering \includegraphics[angle=-90,width=0.8\textwidth]
{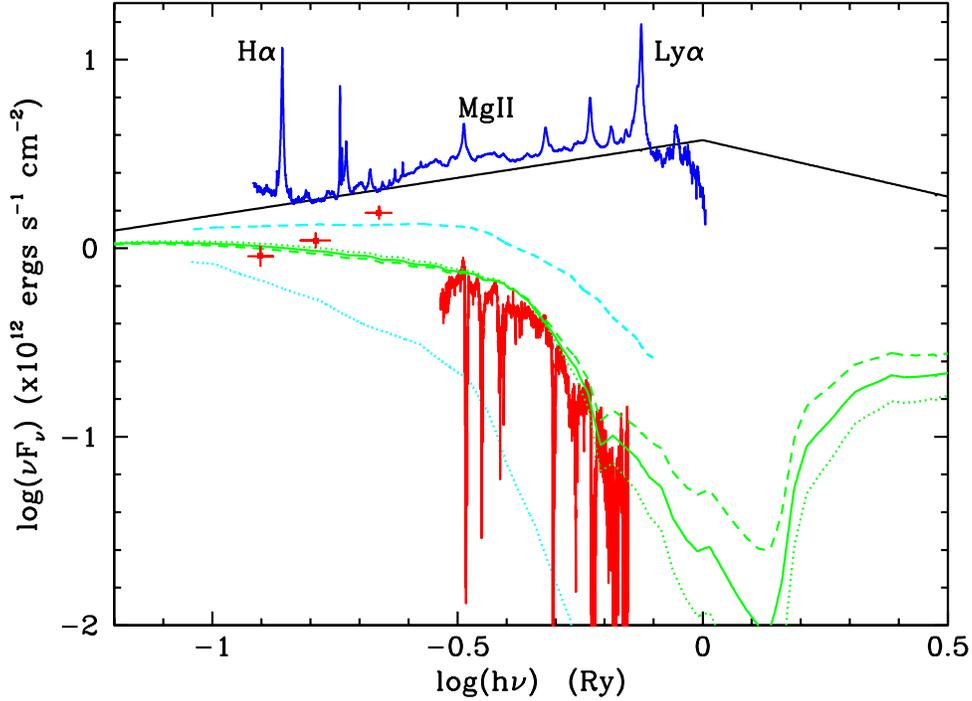}\\
\caption{\footnotesize
Spectrum of SDSS~J0318$-$0600. The observed SDSS spectrum is depicted
by the red curve and the 2MASS J, H, and K photometric measurements
are indicated by red crosses. The black curve depicts our adopted UV-soft
SEDand the blue curve shows the FIRST Bright Quasar Survey composite
spectrum \citep{2001ApJ...546..775B}. The green lines show the reddened
UV-soft SED by our distributed silicate grain model with $\alpha=3.4$
(solid line), 3.3 (dotted line), and 3.5 (dashed line). The cyan curves
depict the UV-soft SED reddened by the SMC extinction curve with
$E(B-V)= 0.1$ (dashed line) and 0.4 (dotted line)}
\label{redningfig}
\end{figure}

Figure \ref{redningfig} shows a portion of the SED centered around the rest 
frame energy range of our observations. Here we show the unattenuated UV-soft
SED described in the previous section and the FIRST Bright Quasar Survey 
composite spectrum of \citet{2001ApJ...546..775B}. The observed spectrum 
of SDSS~J0318$-$0600, also shown in the figure, is clearly affected by
extinction. Five reddened versions of the UV-soft SED are also shown here; 
two of them are a result of the standard extinction curve of the Small 
Magellanic Cloud \citep{1982ApJ...255...70H} with 
$E(B-V)$=0.1 and 0.4, and additional curves generated from {\sc cloudy}
models of pure silicate grains with a distribution of the form
\begin{equation}
%dn_{gr} = C n_H a^{-\alpha} da,\  \  \   20\AA < a < 0.25\mum.
dn_{gr} = C \vy{n}{H} \left( \frac{a}{a_0} \right)^{-\alpha} da, ~~~~  20\AA < a < 0.25\mu m.
\end{equation}
These models were simulated with {\sc cloudy} as the transmitted spectrum
of a cloud with a Galactic dust to gas ratio, $\log \vy{n}{H} (cm^{-3})=4.0$, 
$\log \vy{N}{H} (cm^{-2})=21.8$ and $\log \vy{U}{H}=+2.0$. This high 
ionization parameter was chosen in order
to minimize atomic features in the resulting spectrum and to allow grain 
extinction to dominate the attenuation. A cloud of this number density and 
ionization parameter would be located $\sim 30$~pc from the central source 
roughly with the expected realm of the putative clumpy, dusty torus. The 
dust temperature in this cloud has a maximum value of 400K, which is below 
the sublimation temperature of silicate grains ($\sim 1400$~K for silicates).
See \citet{1995ApJ...450..628P,2008ApJ...685..160N} for a more extensive
discussion of the properties of illuminated dusty tori associated with AGN. 
No graphite grains are included in the models because they lead
to the 2175 \AA\ peak, which is absent in the spectrum. 
On the other hand, a large absorption peak from silicates
at approximately 690\AA\ ($\sim$1.3 Ry) does correlate with the 
observed spectrum. The depth of such a peak is dependent on the size
distribution of grains (i.e. in the value of $\alpha$). 
A problem arises in determining the values of $\alpha$ and total
extinction that are proportional to $C$, because from the UV spectrum
alone no unique solution for both values can be found simultaneously. 
Fortunately, SDSS~J0318$-$0600 was also observed by the Two Micron All Sky Survey
(2MASS). The photometric measurements in the J, H, and K bands 
(rest frame 4162, 5602, and 7277 \AA\ respectively) 
are marked in Fig. \ref{redningfig}. It is important to note here that the 
flux in the J band may be affected by H$\delta$ and other
emission lines in the rest frame of the quasar. 
With this large wavelength range we are able to determine the
magnitude of extinction, while the value of $\alpha$ is
determined by fitting the slope of the SDSS spectrum. 
We find $E(B-V)=0.15$ for an extinction curve with a ratio
of selective to total extinction $R_V=A_V/E(B-V)=4.4$
and $\alpha=3.4$.

The extinction curve of SDSS~J0318$-$0600 with $R_V=4.4$ is quite different
from that of the SMC with $R_V\sim 2$ and the diffuse
Galactic Interstellar Medium (ISM) with 
$R_V=3.1$. The steepness of the present curve is more similar to
that of dense molecular clouds and star forming regions
with $R_V\approx 5$. These large values of $R_V$ are characteristic
of a distribution of grains that favors larger grains than in typical
ISM. This would result if either large grains are allowed to accrete 
in a dense environment shielded from UV radiation or small grains
are efficiently destroyed in a violent event, e.g. shocks. Such may
be the environment of the dusty clouds composing the clumpy tori of AGN.

%\clearpage
%\clearpage

\subsection{Bolometric luminosity and number of
ionizing photons of SDSS~J0318$-$0600}

Once the extinction that affects the observed spectrum has been 
determined we are in the position to compute the bolometric luminosity
of the QSO. To do that we adopt the UV-soft SED shown in 
Figure \ref{redningfig} 
of the previous section and integrate over the whole energy space. 
This gives $L_{Bol}=5.0\times 10^{47}$~ergs s$^{-1}$, assuming 
cosmological parameters of $H_0$=70 km s$^{-1}$ Mpc$^{-1}$, 
$\Omega_m$=0.3, and $\Omega_\Lambda$=0.7.

With the above normalization and our adopted (UV-soft) SED we find a
hydrogen ionizing photon rate of $Q=3.6\times 10^{57}$ photons s$^{-1}$.
However, it is unclear at this time where most of the extinction occurs
with respect to the outflow. If the grain extinction of the SED occurs
interior to the outflow, $Q$ should be determined from our preferred
dust reddened model ($\alpha=3.4$ and $E(B-V)=0.15$). For this case,
the effective $Q=2.1\times 10^{56}$~photons~s$^{-1}$. We will later use
$Q$ in conjunction with the derived values of gas number density and
ionization parameter to estimate the distance of the outflow from the
central source.

\subsection{Photoionization modeling} 

We start by assuming constant total hydrogen density ``clouds'' of
solar abundances from Lodders (2003). The optimization of all our
models is based on the column densities of \siII\ and \siiv. This is
because they have constrained column density measurements from the 
spectrum, expand the whole range of ionization stages observed, and 
they come from the same chemical element, thus their relative strengths 
are independent of the chemical composition of the cloud.

In trying to reproduce the measured ionic column densities we have
constructed a number of photoionization models exploring the
physical conditions and our assumptions. For a
qualitative description of the ionization structure of the cloud
see Korista et al. (2008). It is important in all models that
the average electron density predicted by the model coincide with
that diagnosed from the \siII\ lines in $\S$4.1 within 0.1~dex. For
every model, we
compute the electron density averaged over the \siII\ trough, as:
\begin{equation}
<n_e>_{SiII} = \frac{\int_0^{N_H}N_{SiII}~n_e(N_H)~dN_H}{\int_0^{N_H}N_{SiII}~dN_H}. 
\end{equation}
This value is presented in Table~\ref{cloudytab} together with the adopted
$\vy{n}{H}$.

A summary of the results of different models is presented in
Table~\ref{cloudytab}. Here the first three column densities indicate
the species observed, their measured column densities and respective
uncertainties. The subsequent
column densities present the differences between observations and the results 
of various models as described below. Here we also present the optimal
$U_H$ and $N_H$ determined for every model.

\begin{figure}[!h]
%\rotatebox{-90}{\resizebox{\hsize}{\hsize}
\centering
\includegraphics[angle=-90,width=0.85\textwidth]
%{\plotone{Fig12.eps}}}
  {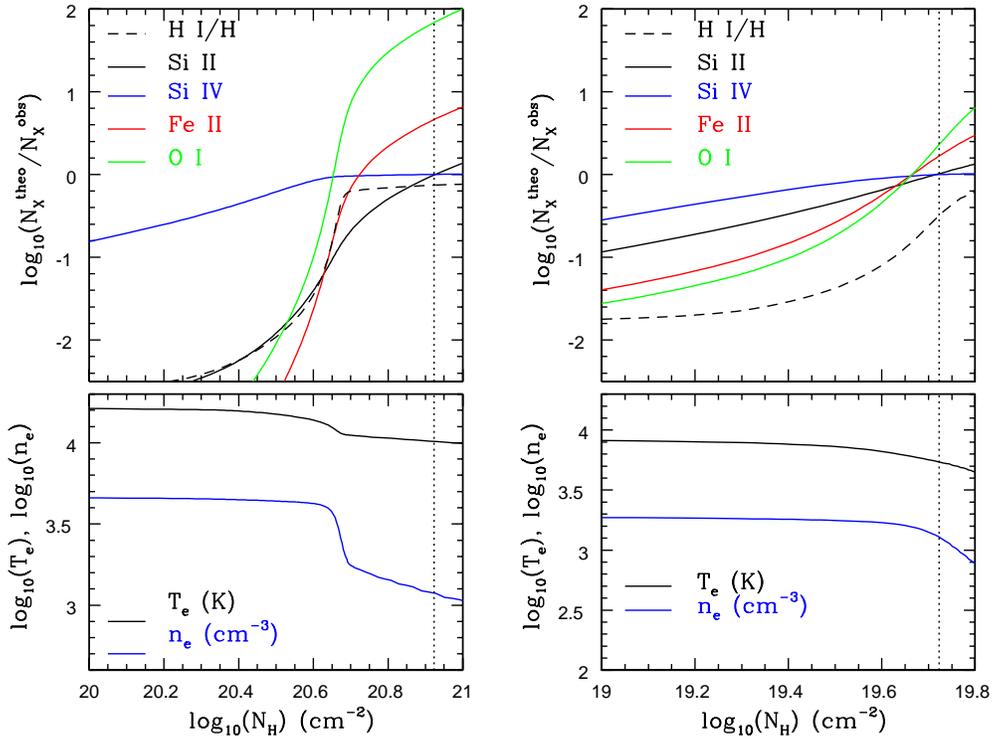}\\
\caption{\footnotesize
Column densities, temperature and electron densities in models of
component {\bf i} of the outflow of QSO~J0318$-$0600. In the upper
panels we show in solid lines the ratios of predicted total
densities of some species to the measured column densities vs.
$\vy{N}{H}$. The dashed lines are neutral hydrogen fractions. The
results of two different models are presented, model B, which uses
solar abundances, a reddened SED and log $U_H$=-2.47, (left panels),
and model F, with the Ballero et al. $Z/Z_\odot=7.2$ abundances, also
with a reddened SED and log $U_H$=-3.13, (right panels). It is found
that when solar abundances are adopted the measured Si column densities
require a total column density that extends beyond the ionization
front, which then results in overestimated \oi\, and \feii\ column
densities}
\label{photofig}
\end{figure}

We first try to model the cloud using solar abundances and two possible
SEDs: our UV-soft SED described above
and the UV-soft SED reddened as in the previous section.
While the measured silicon column densities can be matched in both cases,
this only happens when going very deep into the ionization front where
$n_e$ and temperature have started to drop. Therefore, we see that in
order to reproduce a $\log(<n_e>)=3.3$ in the \siII\ region the particle
density of the cloud must be $\log(<\vy{n}{H}>)=3.8$. Another
concern about these models is their predictions for \oi\
column densities, which come to be about two orders of magnitude
greater than allowed by observations. These results are shown as models 
A and B in Table~\ref{cloudytab} for the unreddened and reddened SEDs 
respectively. Because \oi\ is tightly driven by \hi\ through charge 
exchange, the observed column density of \oi\ is a direct indicator of 
the depth of the ionization front with respect to the total column 
density of the cloud. The same argument is true for the column density 
of \feii. The problems with this model are illustrated in Figure~\ref{photofig}. 
Despite the fact that our measurement of \oi\ is poor the discrepancy between
this and models with solar composition is significant.
The current measured column density is 1.9$\times 10^{15}$~cm$^{-2}$
with an uncertainty of a factor of 2$-$3. On the other hand, models with solar 
composition overestimate the column density by 1.9~dex, i.e. 1.5$\times 10^{17}$~cm$^{-2}$. 
This would be the largest column density of all ions in the spectrum, about 5 times that 
of \civ. Despite the oscillator strength of \oi\ being smaller than that of \civ, 
a column density as predicted would yield a very strong and completely saturated
trough in the spectrum. This is clearly not the case. 
The fact that the present model goes too deep into the ionization front
to reach the high observed ratio of \siII\ to \siiv\ column densities
indicates that the abundance of Si relative to O must be higher than solar.

Similar to Si, the Fe/O and Ni/O abundance ratios must also be supersolar because
the ionic fractions of
\feii, \niII, and \oi\ are tied to neutral hydrogen through charge exchange.
Thus, the \feii/\oi\ and \niII/\oi\ ratios are nearly independent of the physical
condition of the cloud and can only be varied by changing the abundances of these
elements \citep{1998ApJ...492..650B,1996ApJ...460..372B}. Furthermore, we find that
for solar abundances of all other elements, the minimum enhancement of Si, Fe, and
Ni in gas phase that satisfies the are [Si/H]=+0.8 and [Fe/H]=[Ni/H]=+0.3.

We note that this abundances pattern is similar to that expected in high
metalicity bulges of galaxies. \citet{2008A&A...478..335B} studied the evolution
of chemical abundances of spiral bulges hosting Seyfert nuclei by modeling the
efficiency of star formation. They predict the abundances of various elements for
$Z/Z_\odot$ = 4.23, 6.11, and 7.22. We present these abundances, which we use for
photoionization modeling in Table~\ref{abund}.
%\citep[see][Table 2]{2008A&A...478..335B}. 
We present these elemental abundances for the first and last of the metallicities,
and use them in the photoionization modeling to investigate the effects of a
realistic span in gas metallicity. 
\citet{2008A&A...478..335B} do not predict abundances for Al, which is present in
our spectra, thus we simply scale this linearly with $Z$ to [Al/H]=+0.50 and 0.86
at $Z/Z_\odot = 4.23$ and 7.22, respectively. We use the [He/H] vs. $Z$ relation
from those predicted in the chemical evolution models of \citep{1993ApJ...418...11H}.
The remaining elements through Zn untabulated by Ballero et al. were assigned
abundances as estimated by $Z$.

\begin{deluxetable}{lcc|rr|rr|rr} 
\tabletypesize{\scriptsize}
\tablecaption{Measured and model predicted absorption column densities for the
outflow component $\bf i$ in QSO~J0318$-$0600}
\tablewidth{0pt}
\tablehead{
\colhead{} &
\colhead{$\log(N)$ (cm$^{-2}$)}&
\colhead{$\Delta\log(N)$}&
\colhead{$A$} &
\colhead{$B$} &
\colhead{$C$} &
\colhead{$D$} &
\colhead{$E$} &
\colhead{$F$} }

\startdata
$\log(U_H) $                &          &       &  -2.25 &  -2.47 & -2.63 & -3.02 & -2.85 & -3.13 \cr
$\log(N_H) $                &          &       &  20.96 &  20.92 & 20.20 & 20.07 & 19.86 & 19.73 \cr
$\log(\vy{n}{H}) $          &          &       &     3.8&     3.7&    3.3&    3.4&    3.3&    3.3\cr
$<log(n_e)>^{a}$            &          &       &     3.3&     3.3&    3.3&    3.3&    3.3&    3.3\cr
$<log T>_{SiII}$            &          &       &     4.0&     4.1&    3.8&    4.0&    3.9&    3.9\cr
$\log(N_{HI})$              &          &       &  20.50 &  20.43 & 18.50 & 19.36 & 17.48 & 18.26 \cr
Ion  &\multicolumn{2}{c}{Observed} &\multicolumn{6}{c}{$\log(N_X)_{Predicted} - \log(N_X)_{Observed}$}\cr
                & $\log(N_X)$ &$\Delta\log(N_X)$&          &           &           &    &      \cr
\siII     &     16.16&   0.15\tablenotemark{b}&   +0.1 &     0.00& -0.01 &  0.00&  0.00&   0.00\cr
\siiv     &     15.75&   0.11&                    +0.02&     0.00&  0.00 &  0.00& +0.01&  +0.01\cr
\hei*     &   $<$14.0&       &                    -0.1 &    -0.4 & -0.3  & -1.0 & -0.9 &  -1.3 \cr
\cii      &   $>$16.3&       &                    +0.6 &    +0.5 & -0.3  & -0.1 & -0.5 &  -0.5 \cr
\civ      &     16.46&   0.05&                    -0.15&    -0.12& -1.18 & -1.37& -1.43&  -1.62\cr
\oi       &     15.3 & $^{+0.5}_{-0.3}$&          +1.9 &    +1.8 &  0.0  & +0.9 & -0.9 &  +0.3 \cr
\mgi      &   $<$12.3&       &                    +1.5 &    +1.3 & +1.1  & +1.0 & +1.0 &  +1.0 \cr
\mgii     &     15.51&   0.06&                    +0.25&    -0.39& -0.21 & -0.84& -0.29&  -0.66\cr
\alii     &     14.60&   0.04&                    +0.07&    -0.73& -0.10 & -0.89& -0.10&  -0.67\cr
\aliii    &     15.19&   0.06&                    -0.60&    -1.05& -0.73 & -1.15& -0.65&  -1.00\cr
\feii\tablenotemark{c}& 15.38& 0.10&              +0.65&    +0.67& -0.04 & +0.54& -0.69&  +0.21\cr
\niII\tablenotemark{c}& 14.36& 0.30\tablenotemark{b}&+0.45& +0.49& +0.04 & +0.46& -0.57&  +0.23\cr
\znii     &     12.92&   0.02&                    -0.56&    -1.59& -1.26 & -1.39& -1.48&  -1.67\cr
                &          &       &             &          &           &           &   &       \cr
\\
\hline
\enddata
\begin{flushleft}
~~

{\scriptsize \setlength\parindent{0.60in} Model $A$: UV-soft SED and solar abundances;}

{\scriptsize \setlength\parindent{0.60in} Model $B$: Reddened UV-soft SED and solar abundances;}

{\scriptsize \setlength\parindent{0.60in} Model $C$: UV-soft SED and \citet{2008A&A...478..335B} abundances for $Z/Z_\odot=4.2$, see Table~\ref{abund}}

{\scriptsize \setlength\parindent{0.60in} Model $D$: Reddened UV-soft SED and
\citet{2008A&A...478..335B} abundances for $Z/Z_\odot=4.2$, see Table~\ref{abund}}

{\scriptsize \setlength\parindent{0.60in} Model $E$: UV-soft SED and \citet{2008A&A...478..335B} abundances for $Z/Z_\odot=7.2$, see Table~\ref{abund}}

{\scriptsize \setlength\parindent{0.60in} Model $F$: Reddened UV-soft SED and \citet{2008A&A...478..335B} abundances for $Z/Z_\odot=7.2$, see Table~\ref{abund}}

\end{flushleft}
\tablenotetext{a}{Determined from the population ratio of \siII\ excited (287 cm$^{-1}$) to ground levels}
\tablenotetext{b}{Column density error dominated by the uncertainty in oscillator strength}
\tablenotetext{c}{Calculated from theoretical population balance models from the column densities of observed levels}
\label{cloudytab}
\end{deluxetable}
%\clearpage

\begin{deluxetable}{lcc}
\tabletypesize{\footnotesize}
\tablecaption{Abundances used for Photoionization Modeling}
\tablewidth{0.5\textwidth}
\tablecolumns{3}
\noindent \tablehead{
\colhead{Ion} &
\colhead{$Z/Z_\sun$=4.2} &
\colhead{$Z/Z_\sun$=7.2} \\
}
\startdata
$[$Fe/H$]$ & +0.83 & +1.07 \\
$[$O/H$]$  & +0.18 & +0.48 \\
$[$Mg/H$]$ & +0.38 & +0.69 \\
$[$Si/H$]$ & +0.79 & +1.13 \\
$[$Ca/H$]$ & +0.63 & +1.06 \\
$[$C/H$]$  & +0.15 & +0.30 \\
$[$N/H$]$  & +0.28 & +0.87 \\
\enddata
\normalsize
\label{abund}
\end{deluxetable}
\clearpage

The results of these models are given as models C through F in
Table~\ref{cloudytab}. It can be seen that these abundances yield better 
agreement with respect to the measured column densities than when solar 
abundances are used. Moreover, the abundance enhancements for 
$Z/Z_\odot = 4.23$ still yield overestimated column densities for \oi\ 
by one to two sigma, depending on the SED used. On the other hand, the 
models with abundances for $Z/Z_\odot=7.2$ agree well in terms of the 
\oi\ predictions if the reddened UV-soft SED is adopted, but 
underestimate the \oi\ column density when the non-extinguished UV-soft 
SED is used. 

Given our arguments for metalicities greater than solar in the outflow 
and the fact that the solar models are unable to match the $\vy{n}{H}$, 
the true metalicity in QSO~J0318$-$0600 is likely to be several times
solar. Regardless of this we are able to well
constrain $\vy{U}{H}$ and $\vy{N}{H}$. Models C and E constrain 
$\log(\vy{U}{H})$ to $-$2.75 and $-$2.85 and $\log(\vy{N}{H})$ to 
20.16 and 19.86 (cm$^{-2}$), respectively, for the unreddened SED. 
While, models D and F constrain $\log(\vy{U}{H})$ to $-$3.02 and 
$-$3.13 and $\log(\vy{N}{H})$ to 20.07 and 19.73 (cm$^{-2}$), 
respectively, for the reddened SED.

One significant discrepancy between all models and observations is in 
the carbon column densities. As indicated in Table 7, the [Si/C] 
ratio is rather large in these two elemental abundance models, and a 
relatively larger carbon abundance would reduce the discrepancies. 
Ballero et al. point out that their prediction for C is likely 
uncertain as it depends sensitively on the mass of the bulge. However, 
it is important to point out that our measurements for carbon are 
problematic, as we only have a lower limit for the \cii\ column 
density and large portions of the \civ\ troughs are highly saturated.

A relatively small uncertainty in $U_H$ and $N_H$ arises from the adopted 
standard SED. For example, should we adopt the MF87 SED rather than the 
UV-soft SED constructed in Section 4.2, our results for $U_H$ and $N_H$ 
would be affected at a $\sim$0.1 level. However, this effect is very small 
compared to the uncertainties resulting from the dust extinction of the 
SED and the adopted chemical abundances.

Another discrepancy in the models stems from the predictions for Al.
Most of our models underestimate the column densities of \alii\ and 
\aliii. The predicted column density ratios of \alii\ to \aliii\ differ 
from observations by approximately a factor of four. Again, it could be 
that the true Al abundance is higher than 
assumed, which is not surprising as there are only very rough
approximations on how this element scales with metalicity. But, 
beyond the absolute column densities there may also be differences
in the predicted ionization structure of Al. The other problem is that the 
predicted column densities of \mgi\ always exceed the upper 
levels derived from our spectra. It is clear that the ionization 
structure of neutrals that form by {\it recombination} across the ionization 
front is not well described by our models. The cause for this problem 
is a combination of inaccuracies in the atomic parameters for these 
species and limitations of photoionization modeling codes to treat 
neutral ions with low ionization potential ($<$10.2~eV). The exception 
to this is \hei*, which is consistent 
with the models predictions. This problem is intrinsic to all 
modeling codes that treat radiative transfer through the escape 
probability approximation, as neutral species with low ionization 
potential are greatly affected by Lyman~$\alpha$ radiation. Thus, a 
very accurate treatment of radiative transfer is needed to confidently
model these neutral species, which is beyond the scope of this paper.

\subsection{The \niII\ lines as independent distance indicators}

Lucy (1995) demonstrated  that \niII\ could be efficiently photo-excited
by continuum radiation under nebular conditions. Bautista et al.
(1996) constructed a detailed 142-level model of the \niII\ system
for spectral analysis, which was updated with improved atomic data
by Bautista (2004).

\begin{figure}[!h]
%\rotatebox{0}{\resizebox{\hsize}{\hsize}
\centering 
\includegraphics[width=8.5cm,height=13.5cm,angle=0]
%{\plotone{Fig13.eps}}}
{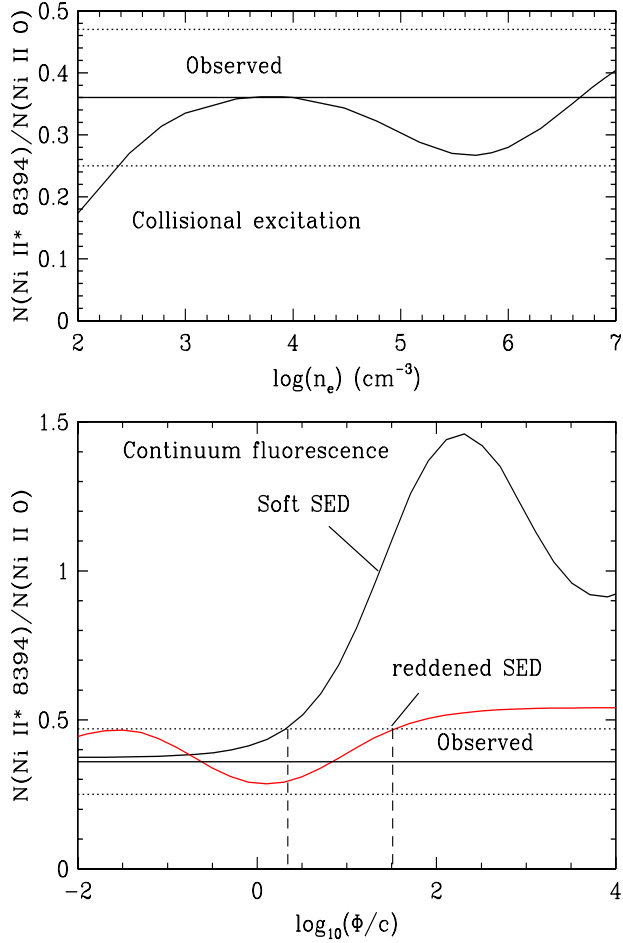}\\
\caption{\footnotesize Ratio of level populations between the \niII\ 
excited level at 3d$^8$4s~$^4$F$_{9/2}$ level at 8394~cm$^{-1}$ and the 
ground level. The ratio is plotted against $\log(n_e)$ for pure  
collisional excitation (upper panel) and $\log(\Phi/c)$ for continuum 
fluorescence excitation. The observed ratio is depicted by the solid 
horizontal line and the uncertainties are shown by dotted lines and the
vertical dashed lines represent the upper limits for $\log_{10}(\Phi/c)$
as constrained by the measured ratio.}
\label{ni2fig}  
\end{figure}
%\clearpage

Figure~\ref{ni2fig} shows the level populations ratio between the excited
level at 8394~cm$^{-1}$ and the ground level. The upper panel shows the 
ratio vs. $n_e$ under pure collisional excitation conditions. 
It is noted that the uncertainty in the measured ratio column
densities is dominated by that in the oscillator strengths, particularly
for troughs from the 8394~cm$^{-1}$. For these transitions we
adopted oscillator strengths from Kurucz and Bell (1995), which in the 
case of the
resonant transitions were discrepant by up to a factor of two from
experimental determinations of Fedchak and Lawler (1999). Thus,
we adopt a nominal 30\% uncertainty in the observed ratios.
Further theoretical and experimental work on the oscillator strengths
of \niII\ is warranted.

From Figure~\ref{ni2fig} one sees 
that the observed ratio is consistent with the
predictions of pure collisional excitation 
for a wide range in $n_e$, that includes the conditions of
the absorbed as diagnosed from \siII.
Hence, it should be possible to set an upper limit to the continuum flux
that also contributes to the excitation of the 8394~cm$^{-1}$ level.
In the lower panel of the figure we plot the ratio against the flux of 
ionizing photons $\Phi$ for the same SED used in our photoionization 
models. It is important to notice that only photons less energetic 
than 1~Ry participate in photoexcitation of the ion and these are 
unaffected by the main atomic opacities of the cloud. It is found that 
the N(\niII\ 8394)/N(\niII\ 0) ratio is strongly enhanced by 
continuum fluorescence and this drives the ratio well above the 
Boltzman equilibrium populations. Under these conditions, induced 
photoexcitation of highly excited odd parity levels is balanced by rapid 
spontaneous radiative decay. For much more intense radiation fields, 
induced photoexcitation exceeds spontaneous decay and the level 
population ratio reaches a plateau. It can be seen that the observed column 
density ratio sets upper limits of 
$\log_{10}(\Phi/c)\approx +1.5$ or +0.4 depending on whether one uses a
reddened or unreddened SED. Thus, from a total particle density of 
$\log(\vy{n}{H})=3.3$ one derives a value for the ionization parameter 
$\log_{10}(U_H)\le -1.8$ or $\le -2.9$, for reddened and unreddened SEDs
respectively. Both of these values are consistent with that derived through 
photoionization modeling. It is unfortunate that more stringent constrains 
could not be derived due to the large uncertainties in the oscillator
strengths for the absorption lines in the spectrum.

\section{Kinetic Luminosity and Mass Flux of the Outflow}

\subsection{Calculating $\dot{M}$ and $\dot{E}_k$ for the Main Outflow Component}

In the previous section, we determined the ionization parameter,
number density and total column column density of the main outflow
component, ($\bf i$). Here we calculate the mass and energetic charateristics
of the outflow. First, we determine the mass contained within the
wind. For simplicity, we assume that  the outflow is in a form of a partial
thin spherical shell at a distance $R$ from the central source moving
radially at constant speed $v$. The total mass contained
in such an outflow is:
\begin{equation}
M = 4 \pi \mu m_p \Omega R^2 N_H,
\end{equation}
where $N_H$ is the total column density of hydrogen, $m_p$ is the mass
of a proton, $\mu$=1.4 is the mean molecular weight of the plasma per
proton, and $\Omega$ is the fraction of the shell occupied by the
outflow. Therefore, assuming d$N_H$/dt=0, the mass flux follows as the
temporal derivative:
\begin{equation}
\dot{M} = 8 \pi \mu m_p \Omega R N_H v,
\end{equation}
the momentum flux is given by:
\begin{equation}
\dot{p}=\dot{M} v,
\end{equation}
and the kinetic luminosity is given by:
\begin{equation}
\dot{E}_k = \frac{\dot{M} v^2}{2} = 4 \pi \mu m_p \Omega R N_H v^3.
\end{equation}

For component $\bf i$: $v=-4200$ km s$^{-1}$ and table 6 gives the
$N_H$ for each of our photoionization models. To calculate the distance
$R$ we use equation (4), where $n_e$ is determined in \S~4.1, $U_H$ is
 taken from table 6 and $Q_H$ is given in \S~4.4, log~$Q_H$ (photons 
s$^{-1}$) as 56.3 and 57.6 photons s$^{-1}$ for the attenuated and 
unattenuated SEDs, respectively. In Table~\ref{quantab} we list the
derived values for $R$, $\dot{E}_k$ and $\dot{M}$ for each 
photoionization model.

\begin{deluxetable}{lc|ccc|ccc}
\tablecaption{Calculated Values for Each Photoionization Model}
%\tablewidth{0.8\textwidth}
\tablecolumns{8}
\tablehead{
\colhead{} &
\colhead{} &
\multicolumn{3}{|c|}{Unattenuated SED} &
\multicolumn{3}{|c}{Attenuated SED} \\
}
\startdata
Abundances             && Solar   & 4.2$^a$ & 7.2$^b$ & Solar   & 4.2$^a$  & 7.2$^b$ \\
Model                  && A       & C       & E       & B       & D        & F     \\
log $U_H$              && $-$2.25 & $-$2.63 & $-$2.85 & $-$2.47 & $-$3.02  & $-$3.13 \\
log $N_H$ (cm$^{-2}$)  && 20.96   & 20.20   & 19.86   & 20.92   & 20.07    & 19.73   \\
$R$ (kpc)              && 9.4     & 14.6    & 18.7    & 2.9     & 5.5      & 6.3     \\
$\dot{E}_k$ ($\Omega_{0.2}$ 10$^{45}$ ergs s$^{-1}$)$^c$&& 12   & 3.1  & 1.8  & 3.3  & 0.9  & 0.5 \\
$\dot{E}_k$/$L_{Bol}$ ($\Omega_{0.2}$)                  && 0.02 & 0.006& 0.004& 0.007& 0.002& 0.001\\
$\dot{M}$ ($\Omega_{0.2}$ M$_\odot$ yr$^{-1}$)          && 2080 & 560  & 330  & 590  & 160  & 80  \\
\enddata
\tablenotetext{a}{Ballero et al. (2008) abundances for $Z/Z_\odot$=4.2}
\tablenotetext{b}{Ballero et al. (2008) abundances for $Z/Z_\odot$=7.2}
\tablenotetext{c}{$\Omega_{0.2}\equiv\Omega$/0.2}
\normalsize
\label{quantab}
\end{deluxetable}
\clearpage

\subsection{The Global Covering Fraction}

As shown above, we determined $\vy{N}{H}$ and $R$ for the photoionization models
 and therefore each model is associated with a reliable estimates of the kinetic
luminosity per unit solid angle. In order to obtain estimates for the
total kinetic luminosity we need to address the remaining variable in
equation (1), $\Omega$, the fraction of the full sphere occupied by
the outflow. There is no direct way to obtain the $\Omega$ of a given
outflow from its spectrum as we only see the material along the line
of sight. Statistically, \civ\ BALs are seen in 20\% of all quasars
\citep{2003AJ....125.1784H}. This can be interpreted as: a) all quasars
have BAL winds covering 20\% of the solid angle, b) 20\% of all
quasars have winds covering the full solid angle, c) all quasars have a
phase of full solid angle wind lasting 20\% of their duty cycle. A
combination of these 3 options is also possible. It is
important to note that with all else equal, options a) and c) yield
the same integrated mass flux and kinetic luminosity over the life
time of an individual quasar, and that all 3 options yield similar
integrated quantities for a sample of quasars.

However, we cannot measure distances to the majority of  \civ\ outflows,
but only to those outflows that show troughs from excited and metastable
levels. For most of our objects (including SDSS~J0318-0600), the spectra
cover such transitions only from singly ionized species (e.g., \feii,
\siII, \cii, and \niII).  Outflows with troughs from these ions are rare
\citep{2006ApJS..165....1T}. Our comprehensive search of the 50,000
brightest SDSS quasars yielded 100 such outflows. We found that only
$\sim$ 1\% of the objects that show a \civ\ outflow show troughs from
these singly ionized species. We therefore have two limits to explore
with regards to the $\Omega$ of the outflow. First, outflows that
exhibit troughs from these 4 singly ionized species excited states may
represent all regular \civ\ outflows viewed through special
conditions that allow for the formation of singly ionized troughs. This
scenario yields $\Omega\simeq0.2$ as is the case for the frequency
of \civ\ BALQSOs among all quasars. The second limit is
$\Omega\simeq0.002$, which is the minimal covering fraction needed to
explain the 100 occurrences in 50,000 objects.

Several arguments suggest that the first scenario is closer to the
actual physical situation. Since outflows with troughs from singly ionized
species are rare, we may simply be observationally biased by this low
ionization selection effect as we lack excited and metastable level
troughs from \civ\ or species with similar ionization. One physically
motivated picture is that we are looking at a normal \civ\ outflow
through a line of sight passing through semi-opaque material at the
edge of the putative AGN torus. Such a scenario explains why we see
the singly ionized species (lowering of the ionization parameter); why
most of the objects we measure have strong dust reddening (torus dust)
and why these systems are so rare (a solid angle that has sufficient
torus material to lower the ionization parameter significantly but not
enough to make it a type II quasar).

Recent observations by Aoki et al. (2009, in preparation) give
support to the above scenario. A quasar showing strong \feii\
absorption troughs and strong dust reddning in 2002, showed much less
dust extinction and very little \feii\ absorption in 2008. Thus, the 
amount of dust along our line of sight changed drastically between two 
epochs seperated by $\sim$3.5 years in the rest frame of the source. The
interpretation is that the dust source has vacated the line-of-sight,
therefore increasing the ionization parameter of the outflow and
reducing the column densitiy of \feii. A 3.5 years time scale favors
changes in a near-by obscuring material (i.e., the putative torus)
over material at a kpc scale and therefore support both of our
assertions: 1) that the dusty material seen in the spectra of these
outflows is indeed much closer to the source than the outflows
themselves; 2) that these excited-state singly ionized absorption
outflows are naturally explained by a special line-of-sight grazing 
the putative torus.

Further support for this scenario comes from the fact that all our
objects show a \civ\ BAL, which is morphologically indistinguishable
from \civ\ BALs in the ubiquitous high ionization BALQSOs. We
therefore conclude that it is more likely that we are seeing normal
$\Omega\simeq0.2$ outflows through a rare line of sight.

A direct resolution of this issue necessitates imaging of the winds.
This is a difficult task due to the high redshifts of the objects and
the presence of a very bright quasar at the center of the image.
Nonetheless, we hope to pursue this line of research in the future.
Another approach is to observe at wavelengths shorwards of \Lya\ where
there are a few
excited state troughs from higher ionization species. In particular
\siv/\siv*$\lambda\lambda$1062.66,1072.97. This ion is similar to
\civ\ in ionization structure and indeed \siv\ troughs are much more
prevalent than those of the singly ionized species. Measuring
$\vy{N}{H}$ and $R$ for these objects will therefore eliminate the
uncertainty connected with the rarity of outflows showing troughs from
singly ionized species, and will yield robust measurements for
$\dot{M}$ and $\dot{E}_k$. We are currently pursuing this project,
which is challenging due to the heavy \Lya\ forest contamination.

\subsection{The Super Massive Black Hole and Host Galaxy}

Given the bolometric luminosity and assuming that the super massive
black hole (SMBH) is accreting near the Eddington limit we find that 
the amount of mass being accreted is:
\begin{equation}
\dot{M}_{acc} = \frac{L_{Bol}}{\epsilon_r c2} = 88~M_{\sun}~yr^{-1},
\end{equation}
where we assumed an accretion efficiency ($\epsilon$$_r$) of 0.1
\citep{1964ApJ...140..796S} and $c$ is the speed of light. This large
accretion rate (the outcome of a very luminous quasar) suggest that 
the mass flux values we derive (see Table 8) are not unreasonable. We
expect such an outflow to have roughly an order of magnitude higher
mass flux than the accretion rate onto the SMBH since the original
outflow entrained material on its journey to several kpc distance.
Simple momentum conservation requires that an outflow which started 
at $\sim$20,000-30,000 \kms to have 5-7 times more mass flux by the 
time it slowed to 4200 \kms. Therefore, an outflow outflow such as
this could have began with $\dot{M}_{outflow}\sim\dot{M}_{acc}$. 

Next, assuming Eddington accretion, we can derive the SMBH mass
and infer a mass for the host galaxy bulge. The black hole mass
derived from Eddington accretion is \citep{1997iagn.book.....P}:
\begin{equation}
M_{BH} = \frac{L_{Bol} \sigma_T}{4 \pi G m_p c}  = 4.0\times10^{9}~M_{\sun},
\end{equation}
where $\sigma$$_T$ is the Thompson scattering cross-section for an
electron, $G$ is the gravitational constant, $M_{BH}$ is the mass of
the SMBH, and $m_p$ is the mass of the proton. Thus, the spheroidal
bulge component of the host galaxy (assumed to be approximately the
mass of the host galaxy) has a mass $M_{bulge}$=3.3$\times$10$^{12}$
M$_{\sun}$ based on a SMBH to bulge ratio of 0.0012
\citep{2002MNRAS.331..795M}.  This mass can be compared with the total
mass ejected by the outflow over the $\sim10^8$ year duty cycle of
the quasar, which is a few times 10$^{10}$ M$_{\sun}$.

\section{Discussion}

\begin{deluxetable}{lcccccc}
\tablecolumns{7}
\tabletypesize{\footnotesize}
\tablecaption{Properties of Measured Outflows to Date}
\tablehead{
\colhead{Object} &
\colhead{$R$$^a$} &
\colhead{log $N_H$} &
\colhead{log U$_H$} &
\colhead{log $\dot{E_k}$} &
\colhead{$\dot{M}$} &
\colhead{Reference$^b$} \\

\colhead{\small{}} &
\colhead{\small{(kpc)}} &
\colhead{\small{(cm$^{-2}$)}} &
\colhead{\small{}} &
\colhead{\small{(ergs s$^{-1}$)}} &
\colhead{\small{($M_\sun$ yr$^{-1}$)}} &
\colhead{\small{}} \\
}
%\footnotesize
\startdata
QSO 0059-2735   & 0.001 - 0.05 & $\gtrsim$21.5$^c$ & $-$0.7      & $\gtrsim$41.1 - 42.8&$\gtrsim$0.2& 1 \\
3C~191          & 28           & 20.3         & $-$2.8           & 44.0         & 310        & 2 \\
QSO 1044+3656   & 0.1 - 2.1    & 20.0 - 22.0  & $-$1.0 - $-$6.0  & 44.5 - 45.4  & 74 - 530   & 3 \\
FIRST 1214+2803 & 0.001 - 0.03 & 21.4 - 22.2  & $-$2.0 - $-$0.7  & 41.6 - 43.8  & 0.3 - 55   & 4 \\
FIRST 0840+3633 & 0.001        & $\sim$21.3   & $<$$-$1.8        & $>$41.9      & $>$0.3     & 5 \\
FIRST 0840+3633$^d$ & 0.23     & $-$          & $-$              & $-$          & $-$        & 5 \\
QSO 2359-1241   & 3            & 20.6         & $-$2.4           & 43.7         & 93         & 6 \\
SDSS J0838+2955 & 3.3          & 20.8         & $-$1.9           & 45.7         & 590        & 7 \\
SDSS J0318-0600 & 6 or 17      & 19.9 or 20.0 & $-$3.1 or $-$2.7 & 44.8 or 45.4 & 120 or 450 & 8 \\
\enddata
\tablenotetext{a}{For relative accuracies, see Section 1.}
\tablenotetext{b}{1-Wampler et al. (1995), 2-Hamann et al. (2001), 3-de Kool et al. (2001),
4-de Kool et al. (2002a), 5-de Kool et al. (2002b), 6-Korista et al. (2008),
7-Moe et al. (2009), 8-This Work}
\tablenotetext{c}{Based on Table 5 in Wampler et al. (1995)}
\tablenotetext{d}{Distance derived from \feii\ fluorescence and no photoionization modeling
was performed for this object}
\normalsize
\label{summary}
\end{deluxetable}

In this section we compare the results from this object to other
outflows found in the literature; discuss the location of the dust in
SDSS~J0318$-$0600, which is a major cause for uncertainty in the
derived mass flux and kinetic luminosity; elaborate on the
relationship between the dynamical time scale of the outflow and the
duty cycle of the quasar; and conclude by describing the possible role
of these outflows in AGN feedback phenomena

In Table \ref{summary} we compare all the analyses in the literature
that can be used to obtain mass flux and kinetic luminosity for
outflows in individual quasars. We also note that with the exception
of SDSS~J0318$-$0600, every absorber shows evidence for inhomogeneous
line of sight (LOS) covering. For outflow absorption, we refer to
homogeneity as the spatial distribution of gas across our line of
sight to the quasar (see section 3.2), where a homogeneous flow would
uniformly cover the background emission source (LOS covering factor of
1). Initially, this may be attributed is to the large
distance of the absorber, however, the outflow in
3C~191 shows evidence of inhomogeneous covering at a greater distance
than we find component {\bf i} of SDSS~J0318$-$0600.

SDSS~J0318-0600 is also the most heavily reddened in the sample, which
leaves us with the question of the location of the obscuring
dust. Based on the value from \citet{1998ApJ...500..525S}, we know
that the Milky Way Galactic reddening is minimal, and thus the
reddening is intrinsic to the quasar.  There exist three possible
locations of the dust with respect to the absorber: interior,
dispersed within, and exterior. The two locations for the dust we
consider here due to their potential impact on the modeling of the
outflow are that the dust is interior to the absorber or the dust is
associated with the absorber. If the dust is interior to the absorber,
a plausible source of dust would be the edge of the putative obscuring
torus of the AGN unified model \citep{1985ApJ...297..621A}. This is
supported by the object discussed by Aoki et al. (2009, in
preperation) and furthermore would provide the rare sight line needed
to match the frequency of detections for low ionization BAL outflows
that show iron absorption (FeLoBAL), as discussed in section 5.1. By
contrast, galactic wind imaging in nearby galaxies shows that dust
co-exists inside the winds and may be associated with the outflows
\citep[e.g., M82,][]{1994ApJ...433..645I}. \citet{2002ApJ...567L.107E}
proposed the possibility of grains forming within AGN outflows, as
conditions allowed. We note though that if the outflowing gas contains
the observed dust, the amount of dust depletion must be substantially
smaller than in the ISM or SMC in order to produce the observe ionic
column densities. Therefore, while we cannot strongly exclude the
possibility of a dusty outflow, locating the dust at the edge of the
putative torus is physically more plausible and explains several
observed features in the FeLoBAL population: the heavy reddening of 
FeLoBALs, the low reddening experienced by \civ\ BAL quasar; and the 
higher prevalence of high ionization BALSQSOs compared with FeLoBALs.

The dynamical time for component $\bf i$ of the outflow is
$t_d=R/v\approx10^6$ yr (assuming $R=6$ kpc). Therefore, if we are
seeing a representative stage of the outflow, over its $\sim$10$^8$ yr
duty cycle \citep []{2005ApJ...625L..71H} the quasar must eject
roughly hundred times more mass than is seen in component $\bf i$. 
Obvious candidates for more ejected mass are the other 10
components of the outflow. However, Bautista et al. (2009, in prep)
show that these components collectively do not contribute
appreciably to the total mass output for SDSS~J0318$-$0600. Earlier 
episodes of mass ejection may not be detectable in
absorption. Let us assume that component $\bf i$ is continuing at
a constant velocity and no longer entraining any significant mass.
Within $10^7$ yr, its distance will increase by a factor of ten and 
therefore its optical depth will drop by a factor of a hundred, making 
it undetectable in absorption. To complete the picture we note that in 
addition to the FeLoBAL system, SDSS~J0318$-$0600 has a high ionization 
BAL at $\sim$27000 km s$^{-1}$. This outflow component is probably 
much closer and therefore represent a very young mass ejection episode 
that in time may evolve to something resembling component $\bf i$. 
Similar situations exist in other objects 
\citep[e.g. SDSS~J0838+2955;][]{2009ApJ...706..525M}.

Given these caveats the kinetic luminosity, such as the one we have
measured in \\
SDSS~J0318$-$0600, provides
$\sim$(0.1-0.6)$\Omega_{0.2}$\% of the bolometric luminosity as
feedback. We note that the measured mass of component $\bf i$
represents only the cold phase (T=10$^4$ K) of the outflow. The hot
phase of the gas (or X-ray, T=10$^6$ K) has been shown in Seyfert
galaxies \citep[e.g., NGC 3783;][]{2003ApJ...599..933N} to contribute
up to 50$\times$ more column density than the cold phase, therefore it
is possible that the gas in the hot phase of the outflow in
SDSS~J0318$-$0600 could contribute significantly to the mass flux and
kinetic luminosity. For comparison, AGN feedback models of galaxy
growth and evolution require a mechanical energy input of roughly 5\%
of the bolometric luminosity of the quasar. For example,
\citet{2005Natur.433..604D} use this ratio to recreate the observed
M$_{BH}$-$\sigma$ relation and \citet{2005ApJ...630..705H} show
in their models that the same value is required to `uncover' a 
shrouded quasar after a merger. Thus, the absorption outflow we observe 
in SDSS~J0318$-$0600 can be a significant contributer for AGN feedback 
processes.

Beyond the effects of feedback on the host galaxy, if SDSS~J0318$-$0600
is representative of the type of quasars found in clusters, then we
also see that an outflow with the kinetic luminosity of
SDSS~J0318$-$0600 would contribute energy to the gas that forms the
ICM. The models of \citet{2004ApJ...608...62S} show that a 5\%
feedback ratio would suffice to heat the ICM during formation of the
cluster and provide the well known L$_x$-T relationship
\citep{1986MNRAS.222..323K,1999MNRAS.305..631A}
Also, related to clusters are X-ray cavities in the ICM, which coincide
with relativistic radio jets \citep[e.g. NGC1275 in the Perseus
cluster,][]{2006MNRAS.366..417F}. We suggest that the sub-relativistic
outflows could be related to the bubbles or perhaps are the driving
mechanisms for bubble inflation. This follows as the inflation of
a bubble is a simpler process provided a wide angle wind 
\citep{2007ApJ...656L...5S}. \citet{2006ApJ...652..216R} shows that the 
power needed to inflate these bubbles is $\sim$10$^{45}$ ergs s$^{-1}$, 
which is similar to the kinetic luminosity we measure in SDSS~J0318$-$0600 
of $\sim$1-3$\times$10$^{45}$ ergs s$^{-1}$. These similarities in fraction 
of the bolometric luminosity and power output suggest that quasar outflows 
play significant roles in influencing their host galaxy and neighborhoods
through feedback.

\section{Summary}

1. We present 6.2 hours of VLT/UVES high resolution spectroscopic 
observations of the BAL quasar SDSS~J0318$-$0600.

2. Using the Al II $\lambda$1671 trough, we divide the outflow into
11 kinematic components ranging in velocity from -7400 to -2800 km 
s$^{-1}$. We utilize these components to identify the corresponding 
absorption troughs in several lines from other ions.

3. For three of these components, we measure many ionic column densities 
and develop methods to estimate ionic column densities in blended and 
saturated regions. The \aliii\ doublet data show that all 11 components 
follow the expected doublet ratio of 2:1 in optical depth and therefore 
the covering of the emission source is homogeneous. Further tests for 
components $\bf i$ and $\bf k$ with \feii\ resonance lines demonstrate 
the same result.

4. The measured column densities of \siII\ and \siII*, allow us to 
determine a number density of 10$^{3.3}$ cm$^{-3}$ for component $\bf i$,
which is the main and innermost component. 

5. We show that the column density ratio of \niII\ excited to resonance 
can be used to independently determine the flux of non-ionizing photons, 
which when combined with the assumed SED yields a direct indication
of the distance. These values agree with that derived from \siII.

6. We analyze the intrinsically reddened spectrum of SDSS~J0318$-$0600, 
and find that it requires a different 
extinction curve from the known Small Magellanic Cloud (SMC) curve. 
We fit the spectrum with a reddened SED due to pure silicate dust and 
a power law grain size distribution with a power of $\alpha$=3.4, 
which resulted in an $R_V$=4.4. Based on our model of the intrinsically
reddened SED, we estimate the bolometric luminosity (log L$_{Bol}$ = 47.7) 
and the number of ionizing photons illuminating the absorber with and 
without dust extinction (log Q = 56.3 or 57.6). 

7. For component $\bf i$, our photoionization models suggest supersolar 
abundances. The output total hydrogen column density and the so called 
ionization parameter for both cases with and without dust extinction are 
$N_H$ = 10$^{19.9}$ or 10$^{20.0}$ cm$^{-1}$ and log U$_H$ = $-$3.1 or 
$-$2.7 respectively. 

8. Given the gas density and ionization parameter, and an estimate of the 
source luminosity, we determine the distance to the 
outflow for attenuated and unattenuated SEDs of 5.9 or 17 kpc respectively, 
and determine the corresponding kinetic luminosities to be (0.7 or
2.5)$\times$10$^{45}$ $\Omega_{0.2}$ ergs s$^{-1}$, and the mass 
fluxes to be (120 or 450) $\Omega_{0.2}$ M$_{\sun}$ yr$^{-1}$. We find that 
component $\bf i$ is the dominant source of energy and mass 
when compared to the other kinematic components, which are evaluated in 
Bautista et al. (2009, in prep).

9. The ratio $\dot{E}_k$/L$_{Bol}$=(0.1-0.6)$\Omega_{0.2}$\% for component 
$\bf i$ in SDSS~J0318$-$0600, makes this outflow a significant contributor 
to AGN feedback mechanisms.

We acknowledge support from NSF grant AST 0507772 and from NASA LTSA grant NAG5-12867.
MAB received partial support for this work from the NASA Award No. NNX09AB99G

%\include{tables/online}

% The bibliography starts here.

\bibliographystyle{apj}

%\bibliography{paper}
\bibliography{ms}

\end{document}